\documentclass[showpacs,aps,prd,floatfix,amsmath,amssymb,nofootinbib]{revtex4}
\usepackage{cleveref}
\usepackage{multirow}
\usepackage{amsmath}
\usepackage{amsfonts}
\usepackage{amssymb}
\usepackage{subfigure}
\usepackage[utf8]{inputenc}
\usepackage{graphicx}
\usepackage{color}
\usepackage{dcolumn}
\usepackage{bm}
\usepackage{float}
\usepackage{footnote} 
\catcode`\@=11
\def\lsim{\mathrel{\mathpalette\@versim<}}
\def\gsim{\mathrel{\mathpalette\@versim>}}
\def\@versim#1#2{\vcenter{\offinterlineskip
\ialign{$\m@th#1\hfil##\hfil$\crcr#2\crcr\sim\crcr } }}
\catcode`\@=12
\def\be{\begin{equation}}
\def\ee{\end{equation}}
\def\bea{\begin{eqnarray}}
\def\eea{\end{eqnarray}}

\graphicspath{{./Diagrams/}}

\definecolor{greenLinks}{rgb}{0,0.6,0}
\definecolor{blueLinks}{rgb}{0,0,0.6}
\definecolor{redLinks}{rgb}{0.6,0,0}
\definecolor{tempText}{rgb}{0.55,0.10,0.67}
\definecolor{eprintLinks}{rgb}{0.4,0.4,0.4}
\definecolor{journalLinks}{rgb}{0.6,0,0}

\usepackage{mathtools,amssymb,amsmath}
\usepackage{color}
\usepackage{xcolor}
\usepackage{ulem}
\usepackage{cancel}
\usepackage{graphicx,epsf,epsfig}
\usepackage{footnote}
\usepackage{lipsum}
\usepackage{multirow}
\usepackage{tabulary}
\usepackage{rotating}
\usepackage{array}
\usepackage{slashed}

\newcommand{\olgafe}[1]{}
\newcommand{\soul}[1]{}

\begin{document}
\title{Lessons from LHC on the LFV Higgs decays $h \to \ell_a \ell_b$ in the Two-Higgs Doublet Models}

\author{\small M. A. Arroyo-Ure\~na }
\email{marco.arroyo@fcfm.buap.mx}
\affiliation{\small Facultad de Ciencias F\'isico-Matem\'aticas}
\affiliation{\small Centro Interdisciplinario de Investigaci\'on y Ense\~nanza de la Ciencia (CIIEC), Benem\'erita Universidad Aut\'onoma de Puebla, C.P. 72570, Puebla, M\'exico.}

\author{\small J. Lorenzo D\'iaz-Cruz }
\email{jldiaz@fcfm.buap.mx }
\affiliation{\small Facultad de Ciencias F\'isico-Matem\'aticas}
\affiliation{\small Centro Interdisciplinario de Investigaci\'on y Ense\~nanza de la Ciencia (CIIEC), Benem\'erita Universidad Aut\'onoma de Puebla, C.P. 72570, Puebla, M\'exico.}

\author{\small O. F\'elix-Beltr\'an }
\email{olga.felix@correo.buap.mx }
\affiliation{\small Facultad de Ciencias de la Electr\'onica}
\affiliation{\small Centro Internacional de F\'isica Fundamental (CIFFU), Benem\'erita Universidad Aut\'onoma de Puebla, C.P. 72570, Puebla, M\'exico.}

\author{\small M. Zeleny-Mora}
\email{moises.zeleny@cuautitlan.unam.mx}
\affiliation{\small Departamento de F\'isica, FES-Cuautitl\'an,
	Universidad Nacional Aut\'onoma de M\'exico, C.P. 54770, Estado de M\'exico, M\'exico.}

\begin{abstract}
The non-conservation of the lepton number has been explored at the LHC through the Lepton-Flavor Violating (LFV) Higgs decays $h\to\ell_a\ell_b$, with $\ell_{a,\,b}=e,\,\mu,\,\tau$ $(a \neq b)$. Current limits on these decays are a source of valuable information on the structure of the Yukawa and Higgs sectors. The LFV Higgs couplings can arise within the general Two-Higgs Doublet Model (2HDM); the predicted rates for these decay modes depend on the specific Yukawa structure being considered, ranging from a vanishing branching ratio at tree-level for some versions (2HDM-I, II, X, Y), up to large and detectable ratios within the general 2HDM-III. An attractive scenario is given by the texturized version of the model (2HDM-Tx), with the Yukawa matrices having some texture zeros, such as the minimal version with the so-called Cheng-Sher ansazt.  We study the constraints on the parameter space of the 2HDM provided by experimental and theoretical restrictions, and use them to study the detection of LFV Higgs modes at LHC. We find several encouraging scenarios to the search for the decay $h \to\tau\mu$ that could be achieved in the High-Luminosity LHC. On the other hand, LFV Higgs couplings can also be induced at one-loop level in the 2HDM with neutrino masses, with the loops being mediated by neutrino interactions; we find that the resulting branching ratios are of order $10^{-7}$ at best, which is out of the reach of current and future phases of the LHC.
\end{abstract}

\maketitle

\newpage


\newpage

\section{Introduction}\label{section:introduction}
The Standard Model (SM), with a single Higgs doublet generating all the masses of the model, it has been confirmed at the Large Hadron Collider (LHC) thanks to the detection of a light Higgs  boson with $m_h=125$ GeV~\cite{CMS:2012qbp, ATLAS:2012yve}. 
The LHC program includes  measuring the Higgs properties, to determine whether they are consistent with the SM predictions, or show some deviations that could point into the direction of physics beyond the SM~\cite{ATLAS:2016neq}. 
So far, the LHC has been able to measure the Higgs couplings with $WW, \, ZZ$, and 
$b\bar{b}, \tau^+ \tau^-,  \mu^+ \mu^-$, and  indirectly the coupling with top quarks (from the Higgs coupling with photon and
gluon pairs, which arise at one-loop). 

On the other hand,   the open problems of the SM, together with the results in areas such as neutrino physics and dark matter, suggest that some form of new physics should exist,  to account for such phenomena.
So far, a variety of experimental searches, including  the energy, precision and cosmological frontiers,  have
 imposed strong limits on the scale of new physics above $\mathcal{O}$(1) TeV~\cite{Murayama:2007ek}.

Models of new physics often include a Higgs sector with an extended spectrum and some new features~\cite{Branco:2011iw}.  
Bounds on the flavor conserving Higgs couplings that arise in these models  have been derived at LHC too~\cite{Arroyo-Urena:2020fkt}.  Moreover, some of these extensions of the SM, include a more flavored Higgs sector~\cite{Diaz-Cruz:2002ezb},  which produces distinctive signals, such as the Flavor-Violating (FV) Higgs-fermions interactions; and in some cases the 
rates could be so large, that  some suppression mechanism is required in order to build viable models~\cite{LorenzoDiaz-Cruz:2019imm}.

In multi-Higgs doublet models, the Glashow-Weiberg theorem~\cite{{PhysRevD.15.1958}} states that in order to avoid the presence of FV
couplings in the Higgs-Yukawa sector, each fermion type (up- and down-type quarks) must acquire their masses from 
a unique Higgs  doublet. Later on, it was found that it is possible to build realistic models with FV Higgs couplings, 
provided that the Yukawa matrices have a restricted form, {\it i.e.} with texture zeroes~\cite{Cheng:1987rs}. The extension of these ideas to
the lepton sector was presented in Refs.~\cite{Sher:1992kj,Diaz-Cruz:1999sns}.
 
It is possible to probe these Lepton Flavor-Violating (LFV) effects at low-energies, through the  LFV decays $\ell_i \to \ell_j \gamma$, $\ell_i \to \ell_j \ell_j \ell_k$. 
Furthermore, Ref.~\cite{Diaz-Cruz:1999sns, Chattopadhyay:2019ycs, Altmannshofer:2016oaq}  also proved that it is possible to have large rates for the LFV Higgs decays 
and to satisfy those low-energy LFV bounds. 
More realistic textures were considered in~\cite{Diaz-Cruz:2004wsi},
while the  possibility  to search for these LFV Higgs decays 
at hadron colliders was studied in Ref.~\cite{Han:2000jz}.  
 The most recent search for LFV Higgs Decays at LHC,  with center-of-mass energy $\sqrt{s} =13$ TeV 
 and an integrated luminosity  of $138$ fb$^{-1}$, has provided stronger bounds, 
 in particular ATLAS reports both
$\mathcal{BR}(h \to e \tau)<0.20\%$ and $\mathcal{BR}(h \to \tau\mu)<0.18\%$~\cite{ATLAS:2023mvd},  while the CMS collaboration concludes that $\mathcal{BR}(h \to e \tau)<0.22\%$ and ${\cal{B}}(h \to  \tau\mu)<0.15\%$ for $137$ fb$^{-1}$ of accumulated data~\cite{CMS:2021rsq}, which
are consistent with a zero value. The fact that the LHC has searched for these LFV Higgs boson decays modes, and has already presented strong bounds on the corresponding branching ratios,  has motivated great interest from the theoretical side.
Other models where the LFV Higgs signal is present  are models with new physics, including: 2HDM~\cite{Tsumura:2005mv, Kanemura:2005hr, Primulando_2020, Vicente:2019ykr, deGiorgi:2023wjh}, models with a low-scale flavon mixing with the SM-like Higgs boson~\cite{Arroyo-Urena:2018mvl}, models with a gauged $L_\mu - L_\tau$ symmetry~\cite{Altmannshofer:2016oaq}, models with 3-3-1 gauge symmetry~\cite{Hue:2015fbb}; See-Saw  model and its inverse version~\cite{PILAFTSIS199268,PhysRevD.47.1080,PhysRevD.71.035011,THAO2017159}, low scale See-Saw~\cite{PhysRevD.91.015001, PhysRevD.102.113006}, as well as  SUSY models, including models with neutrino masses, such as See-Saw MSSM~\cite{Diaz-Cruz:2008agf} and  the minimal SUSY SM (MSSM)~\cite{BRIGNOLE2003217,DiazCruz:2008ry} at loop-level. Several type of methods has been used to calculate the LFV Higgs decays, for example the Mass Insertion Approximation (MIA)~\cite{Arganda2016,PhysRevD.95.095029,10.3389/fphy.2019.00228} or Effective Field approach~\cite{PhysRevD.99.095040, Alonso-Gonzalez:2021tpo}, and  the minimal SUSY SM (MSSM)~\cite{Diaz-Cruz:2002ezb}. At the coming phase (with higher luminosity) it will be possible to derive more restrictive bounds,  and
one can analyze the consequences for the model building side. 

In this paper we present a general discussion of the implications of the LFV Higgs decays searches at LHC,
within the general 2HDM of type III (2HDM-III)~\cite{LorenzoDiaz-Cruz:2019imm}. We start by study in the limits  on $\mathcal{BR}(h \to \tau \mu)$ that can be achieved at LHC, as a function of the integrated luminosities that is projected for the future phases of the LHC.  Then, we shall derive constraints on the model parameters, and discuss the lessons that LHC can impose on the model structure, and see how the resulting rate fit into the picture. We also discuss the expected rates for the LFV decay $h \to \ell_a \ell_b$, within the $\nu$2HDM, where such coupling is induced at one-loop level.

The organization of our paper goes as follows. Section~\ref{section:thelfvhiggscoup} contains our parametrization for the Higgs couplings for the general
2HDM-III. Section~\ref{section:thelhcsearch} contains the study of the 2HDM-III parameter space which was obtained from low-energies processes and Higgs boson data. Results for the search for LFV Higgs decays at LHC and their future stages, namely, HL-LHC, HE-LHC and FCC-hh, are also presented in Sec.~\ref{section:thelhcsearch}.
Section~\ref{section:oneloop} is dedicated to the one-loop calculation for the decay $h\to \ell_a \ell_b$ within the 2HDM of type I, II,
including massive neutrinos, which are generated using the See-Saw mechanism of type I.  
Finally, conclusions and perspectives are presented in Section~\ref{section:conclusions}. 

\section{The LFV Higgs couplings in the 2HDM's}\label{section:thelfvhiggscoup}

The general two-Higgs doublet model admits the possibility to have LFV Higgs couplings, which have already been explored at LHC through the search for the decays $h \to \ell_a \ell_b$.  The predicted rates for these decays depend on the specific model realization, ranging from a vanishing branching ratio at tree-level for 2HDM-I, II, X, Y, as well as the MFV implementation, up to large and detectable ratios within the version 2HDM-III, which depend on the specific Yukawa structure being considered. For instance, the texturized version of the 2HDM (2HDM-Tx) with Yukawa matrices having for texture zeros~\cite{Cheng:1987rs,Zhou:2003kd,Diaz-Cruz:2004wsi,Diaz-Cruz:2004hrt}, which has been found to reproduce the Cheng-Sher ansazt~\cite{Arroyo-Urena:2013cyf,CarcamoHernandez:2006iuj}, provides the largest possible rates and it is strong by constrained at LHC (Run2), as we shall argue here. 
Next in importance could be the 2HDM where the LFV Higgs couplings are absent at tree level, that could be induced at one-loop, as it is the case for the $\nu$2HDM, which is the version of the 2HDM I, II but includes neutrino masses.

Thus, we start by presenting the coupling of the Higgs boson ($h$) with vector bosons ($W^\pm, \, Z$). In this case 
we can write the interaction Lagrangian with terms of 4-dimension, consistent with Lorentz symmetry and derivable from a renormalizable model, as follows:
\begin{equation}
 {\cal{L}}_V = \kappa_W gm_W h W^{+\mu} W^-_{\mu} + \kappa_Z \dfrac{gm_Z}{2c_W} h Z^{\mu} Z_{\mu}.
\end{equation}
where $g$ is the SU(2) gauge constant, $h$ is the Higgs particle; $m_W$, $m_Z$ are the $W^{\pm}$ and $Z$ gauge bosons masses respectively; and $c_W=\cos \theta_W$.
When the Higgs particle corresponds to the minimal SM, one has $\kappa_W=\kappa_Z=  \kappa_V = 1$, while values 
$\kappa_W=\kappa_Z \neq 1$ arise in models with 
several Higgs doubles, which respect the so-called custodial symmetry. In particular for models
that include $N$ Higgs doublets $\Phi_a$, each one having a vacuum expectation value (vev) $v_a$, one has:
\begin{equation}
 \kappa_V=  \sum_a \dfrac{v_a}{v} O_{1a},
\end{equation}
where $O_{1a}$ denotes the element of the matrix that diagonalizes the $\mathcal{CP}$-even Higgs mass matrix, with $h_1=h$ denoting
the light SM-like Higgs boson, and $v^2=\sum_{a=1}^N v_{a}^2=246^2$ \rm{GeV$^2$}. In the particular case of the 2HDM-III, $O_{11} = O_{12} = \sin(\alpha - \beta)$.

In the Yukawa sector of the 2HDM-III, the mass matrix of fermions is given by:
\begin{equation}
	M^f = \dfrac{v_1}{\sqrt{2}}Y_1^f + \dfrac{v_2}{\sqrt{2}}Y_2^f,
\end{equation}
with $Y^f_a$ ($a=1,2$) are the Yukawa matrices. $M^f$  is diagonalized by means of the unitary matrix $V_f$ through a unitary transformation, so, in the mass basis~\cite{Felix-Beltran:2013tra},
\begin{equation}
	\bar{M}^f = \dfrac{v_1}{\sqrt{2}}\tilde{Y_1}^f + \dfrac{v_2}{\sqrt{2}}\tilde{Y_2}^f, \text{ with } \tilde{Y_a}^f = V_f^\dagger Y_a^f V_f.
\end{equation}
As a consequence, we have,
\begin{equation}\label{ec:y1f_interms_y2f}
	\tilde{Y_1}^f = \dfrac{\sqrt{2}}{v_1} \tilde{M}^f - \tan \beta \tilde{Y_2}^f.
\end{equation}
On the other hand, the interaction of the neutral Higgs $\phi$ with the fermion $f$, with $f=q$ or $\ell$ for quarks and leptons
respectively, can also be written in terms of a 4-dimension Lagrangian that respects Lorentz invariance, namely:
\begin{equation}\label{ec:hll_SP}
 {\cal{L}}_f^\phi =  \phi \bar{f}_i ( S_{ij}^\phi + i \gamma_5 P_{ij}^\phi) f_j,  \, (i,j=1,2,3).
\end{equation}

The $\mathcal{CP}$-conserving ($\mathcal{CP}$C) and $\mathcal{CP}$-violating ($\mathcal{CP}$V) factors $S_{ij}^\phi$ and $P_{ij}^\phi$, which include the flavor physics we are interested in, are written as:
\begin{eqnarray}
 S_{ij}^\phi &=& \dfrac{g m_i  }{2 m_W} c_{f}^\phi \delta_{ij} + \dfrac{g  }{2} d_{f}^\phi \eta_{ij}^\phi, \\
 P_{ij}^\phi &=& \dfrac{g m_i  }{2 m_W} e_{f}^\phi \delta_{ij} + \dfrac{g  }{2} g_{f}^\phi {\eta'}_{ij}^\phi.
\end{eqnarray}

The factors $c_f^\phi, d_f^\phi , e_f^\phi$ and $ g_f^\phi $ depend on new physics in the Higgs sector, where $\phi$ is part 
of an enlarged Higgs sector. They also describe the possibility that the fermion masses may come from more 
than one Higgs doublet.  As far the factors $\eta_{ij}^\phi$, ${\eta'}_{ij}^\phi$ are concerned, they represent new physics associated with
the origin of flavor, {\it i.e.}, they have an explicit dependence  on the Yukawa structure.

Within the SM $c_f^h=1$ and $d_f^h = e_f^h = g_f^h =0$, which signals the fact that for the SM 
the Higgs-fermion couplings are $\mathcal{CP}$C and  flavor diagonal, {\it i.e.}, there is no Flavor Changing Scalar Interactions (FCSI).  
Notice that when we have $e_f^\phi \neq 0$, but $d_f^\phi= g_f^\phi =0$, it indicates
that the Higgs-fermion couplings are $\mathcal{CP}$V, but still flavor diagonal.
In models with two or more Higgs doublets, it is possible to have both FCSI ($d_f^\phi,\, g_f^\phi  \neq 0 $)  and $\mathcal{CP}$V.

Explicit values for the coefficients $c_f^\phi, d_f^\phi$ within the general Two-Higgs doublet model with Yukawa matrix of texture form (2HDM-Tx), are shown in Table~\ref{table-0}~\cite{Arroyo-Urena:2020mgg}, for the $\mathcal{CPC}$ case: $e_f^\phi= g_f^\phi=0$, while the parameters $\eta_{ij}^\phi$ are written as follows:
\begin{equation}
\eta_{ij}^\phi = \dfrac{ \sqrt{m_i m_j} }{m_W} \chi_{ij}^\phi,
\end{equation}
where the coefficients $\chi_{ij}$ parametrize the dependence on the flavor structure  of the Yukawa 
matrices, and it  is expected to be $\mathcal{O}(1)$ in minimal scenarios (such as models with four-texture zeroes that reproduces
the so called Cheng-Sher ansazt). However, this parameter  could be more suppressed in other settings, such as the so-called
Texturized models~\cite{Arroyo-Urena:2019lzv}. In the BGL models~\cite{Branco_1996} it is shown that the size of the FV Higgs couplings is
given by the CKM matrix elements, and therefore the resulting branching ratios could be more suppressed. 

For the charged scalar bosons, we have
%
\begin{eqnarray}\notag
	\mathcal{L}_Y^{c h} 
	& =\left[\bar{d}_i\left(\dfrac{\sqrt{2}}{v}\tan \beta  \bar{M}_{ij}^u \delta_{ij} - \sec\beta\tilde{Y}_{2,ij}^u \right) u_j H^{-}V_{\rm CKM}^{ij*}\right. \\ \notag
	& \left.-\bar{u}_i\left(\dfrac{\sqrt{2}}{v}\tan \beta \bar{M}_{ij}^d   \delta_{ij} - \sec\beta \tilde{Y}_{2,ij}^d \right) d_j H^{+}V_{\rm CKM}^{ij}\right] P_R \\
	& -\left[\bar{d}_i\left(\dfrac{\sqrt{2}}{v}\tan \beta  \bar{M}_{ij}^d \delta_{ij} - \sec\beta \tilde{Y}_{2,ij}^d \right) u_j H^{-}V_{\rm CKM}^{ij*}\right. \\\notag
	& \left.+\bar{u}_i\left(\dfrac{\sqrt{2}}{v}\tan \beta  \bar{M}_{ij}^u \delta_{ij} - \sec\beta\tilde{Y}_{2,ij}^u\right) d_j H^{+}V_{\rm CKM}^{ij}\right] P_L.
\end{eqnarray}
%
%
\begin{center}
	\begin{table}
		\begin{tabular}{|c|c|c|c|c|c|c|}
			\hline 
			Coefficient & $c_{f}^{h}$ & $c_{f}^{A}$ & $c_{f}^{H}$ & $d_{f}^{h}$ & $d_{f}^{A}$ & $d_{f}^{H}$\tabularnewline
			\hline 
			\hline 
			$d$-type & $-\dfrac{\sin\alpha}{\cos\beta}$ & $-\tan\beta$ & $\dfrac{\cos\alpha}{\sin\beta}$ & $\dfrac{\cos(\alpha-\beta)}{\cos\beta}$ & $\csc\beta$ & $\dfrac{\sin(\alpha-\beta)}{\cos\beta}$\tabularnewline
			\hline 
			$u$-type & $\dfrac{\cos\alpha}{\sin\beta}$ & $-\cot\beta$ & $\dfrac{\sin\alpha}{\sin\beta}$ & $-\dfrac{\cos(\alpha-\beta)}{\sin\beta}$ & $\sec\beta$ & $\dfrac{\sin(\alpha-\beta)}{\sin\beta}$\tabularnewline
			\hline 
			leptons $\ell$ & $-\dfrac{\sin\alpha}{\cos\beta}$ & $-\tan\beta$ & $\dfrac{\cos\alpha}{\sin\beta}$ & $\dfrac{\cos(\alpha-\beta)}{\cos\beta}$ & $\csc\beta$ & $\dfrac{\sin(\alpha-\beta)}{\cos\beta}$\tabularnewline
			\hline 
		\end{tabular}
			\caption{Coefficients for $\phi$-Fermion couplings in 2HDM-Tx with $\mathcal{CP}$-conserving Higgs potential.}	\label{table-0}		
	\end{table}
\end{center}
%

\section{Search for the $h\to\tau\mu$ decay at future hadron colliders}\label{section:thelhcsearch}
Let us start with the analysis of the 2HDM-III parameter space. It is worth to mentioning that one of the authors has already analyzed several experimental constrains in Ref.~\cite{Arroyo-Urena:2020mgg} for the theoretical framework presented here. However, we update the model parameter space considering the most up-to-date experimental data from LHC and low-energy processes, as shown below.

\subsection{Constraint on the 2HDM-III parameter space}\label{subsection:constraintonthethdmiii}
\begin{figure}[!ht]
\center{\includegraphics[scale=0.6]{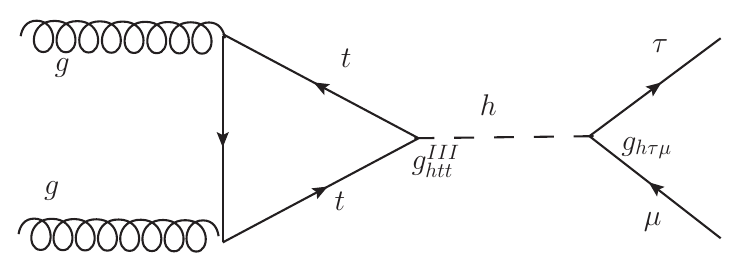}}
 \caption{Feynman diagram of the dominant Higgs boson production at LHC with its subsequent FCSI decay into $\tau\mu$ pair. }
 \label{FeynmanDiagram}
\end{figure}

\begin{figure}[!ht]
	\center{\includegraphics[scale=0.22]{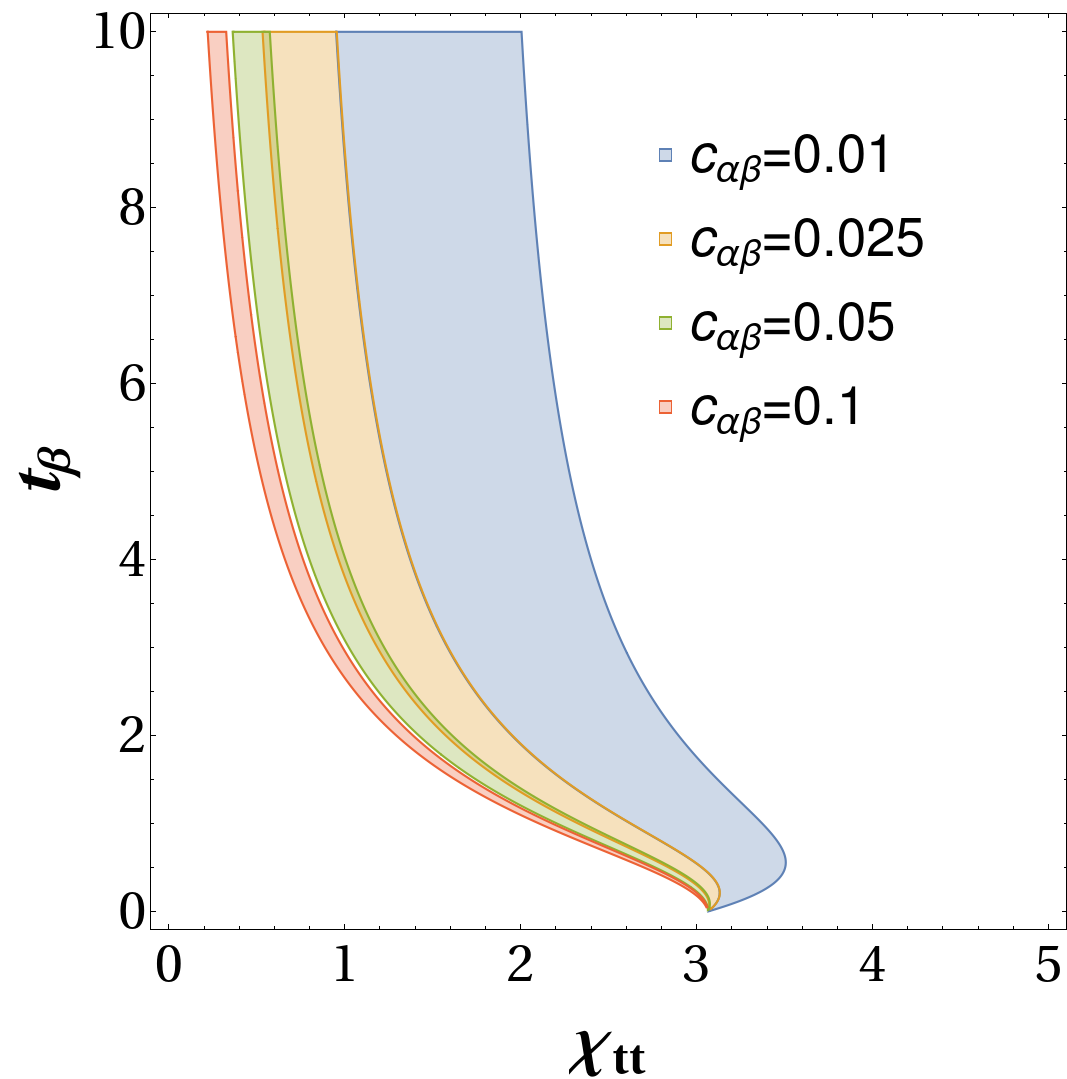}}
	\caption{The colored areas in the $\chi_{tt}-t_{\beta}$ plane represent points agree with the $htt$ SM coupling including uncertainties coming from the top quark mass.}
	\label{tb_chitt}
\end{figure}

To evaluate the production cross section of the SM-like Higgs boson and the branching ratio of the $h \to \tau \mu$ decay, it is necessary to analyze the 2HDM-III parameter space. The relevant 2HDM-III parameters that have a direct impact in the predictions are: $c_{\alpha \beta} \equiv \cos(\alpha-\beta)$ and $t_{\beta} \equiv \tan \beta$. This is so because both $g_{h \tau \mu}$ and $g_{htt}^{III}$ couplings are proportional to them (see Figure~\ref{FeynmanDiagram}), namely,
\begin{eqnarray}
 g_{h \tau \mu}&=&\dfrac{c_{\alpha \beta} t_\beta}{\sqrt{2} s_\beta}\dfrac{m_\mu m_{\tau}}{m_W}\chi_{\tau\mu},  \\
 g_{htt}^{III}&=&\dfrac{gm_t}{2m_W}\Bigg(\dfrac{c_\alpha}{s_\beta}-\dfrac{c_{\alpha\beta}}{s_\beta}\chi_{tt}\Bigg),
\end{eqnarray}
where $s_{\beta}\equiv \cos \beta$, and $m_\mu, \, m_\tau, \, m_t$ are the masses of muon, tau and top respectively. Here we have considered $\chi_{tt}=1$, this value is in accordance with the $htt$ SM coupling (which depends on $t_\beta$ and $c_{\alpha\beta}$, as shown in Figure~\ref{tb_chitt}. In this analysis we include uncertainties coming from direct measurements of the top quark mass~\cite{ParticleDataGroup:2022pth}.    

We use experimental data in order to constrain the free parameters previously mentioned. The software used for this proposal is a  \texttt{Mathematica} package called \texttt{SpaceMath}~\cite{Arroyo-Urena:2020qup}. The physical observables used to bind the free model parameters are listed below: 
\begin{enumerate}
\item LHC Higgs boson data.
\item $B_s^0\to\mu^-\mu^+$ decay~\cite{ParticleDataGroup:2022pth}.
\item Upper limits on the $\mathcal{BR}(\tau\to 3\mu)$, $\mathcal{BR}(\tau\to\mu\gamma$) and $\mu\to e\gamma$ decays~\cite{ParticleDataGroup:2022pth}.
\item Upper limit on the $\mathcal{BR}(h\to\tau\mu)$ decay~\cite{CMS:2017con,ATLAS:2019icc}.
\item The muon anomalous magnetic dipole moment $\delta a_{\mu}$ and the upper limit on the electric dipole moment of the electron $a_e$~\cite{Muong-2:2021ojo, Roussy:2022cmp}.
\item Direct searches for additional heavy neutral $\mathcal{CP}$-even and $\mathcal{CP}$-odd scalars through $gb\to\phi\to\tau\tau$ ($\phi=\,H,\,A$) process are used to constrain their masses~\cite{ATLAS:2017nxi,CMS:2018rmh}. We denote them as $m_H$ and $m_A$ respectively.
\item As far as the charged Higgs scalar mass $m_{H^{\pm}}$ is concerned, the observables to constrain $m_H^{\pm}$ are: {\it (i)} the upper limit on $\sigma(pp\to tbH^{\pm})\times \mathcal{BR}(H^{\pm}\to\tau^{\pm}\nu)$~\cite{ATLAS:2018gfm}, and {\it (ii)} the decay $b\to s\gamma$~\cite{Chetyrkin:1996vx, Adel:1993ah, Ali:1995bi, Greub:1996tg, Ciuchini:1997xe, Misiak:2017bgg}.  
\end{enumerate}
%

\subsubsection{Constraint on $c_{\alpha\beta}$ and $t_{\beta}$}\label{subsubsection:Constrain_Cba-tb}
We first analyze the impact of LHC Higgs boson data on $c_{\alpha\beta}$ and $t_{\beta}$. More precisely, we use the coupling modifiers $\kappa$-factors reported by ATLAS and CMS collaborations~\cite{ATLAS:2018doi,CMS:2018uag}. In the context of 2HDM-III, this observable is defined as follows
\begin{equation}\label{kappa}
\kappa_{pp}^2=\dfrac{\sigma(pp\to h^{\text{2HDM-III}})}{\sigma(pp\to h^{\text{SM}})}\;\text{or}\;\kappa_{x\bar{x}}^2=\dfrac{\Gamma(h^{\text{2HDM-III}}\to X)}{\Gamma({h^\text{SM}}\to X)}.
\end{equation}

\begin{figure}[!ht]
\centering
\subfigure[]{{\includegraphics[scale=0.17]{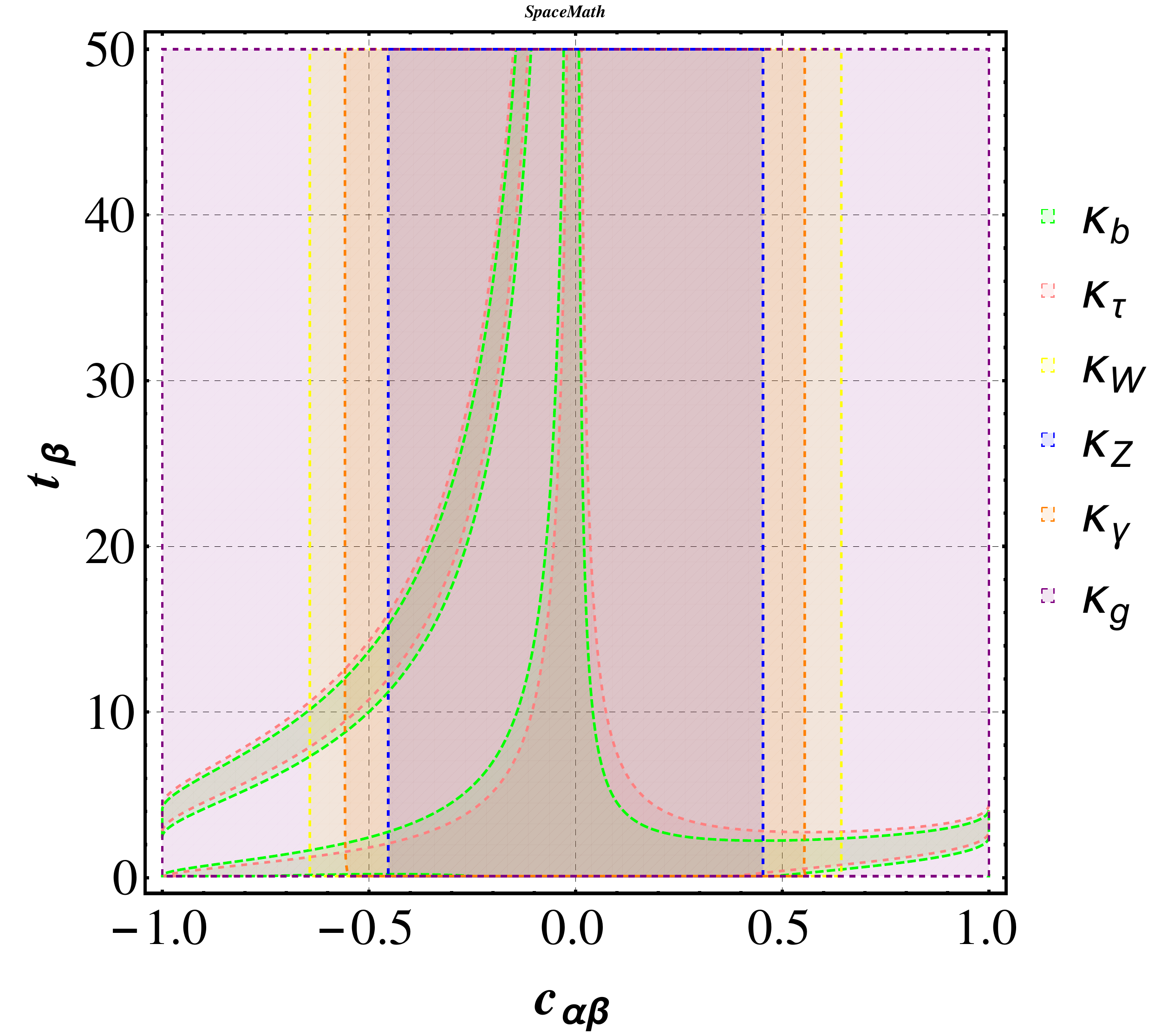}}\label{2a}}
\subfigure[]{{\includegraphics[scale=0.27]{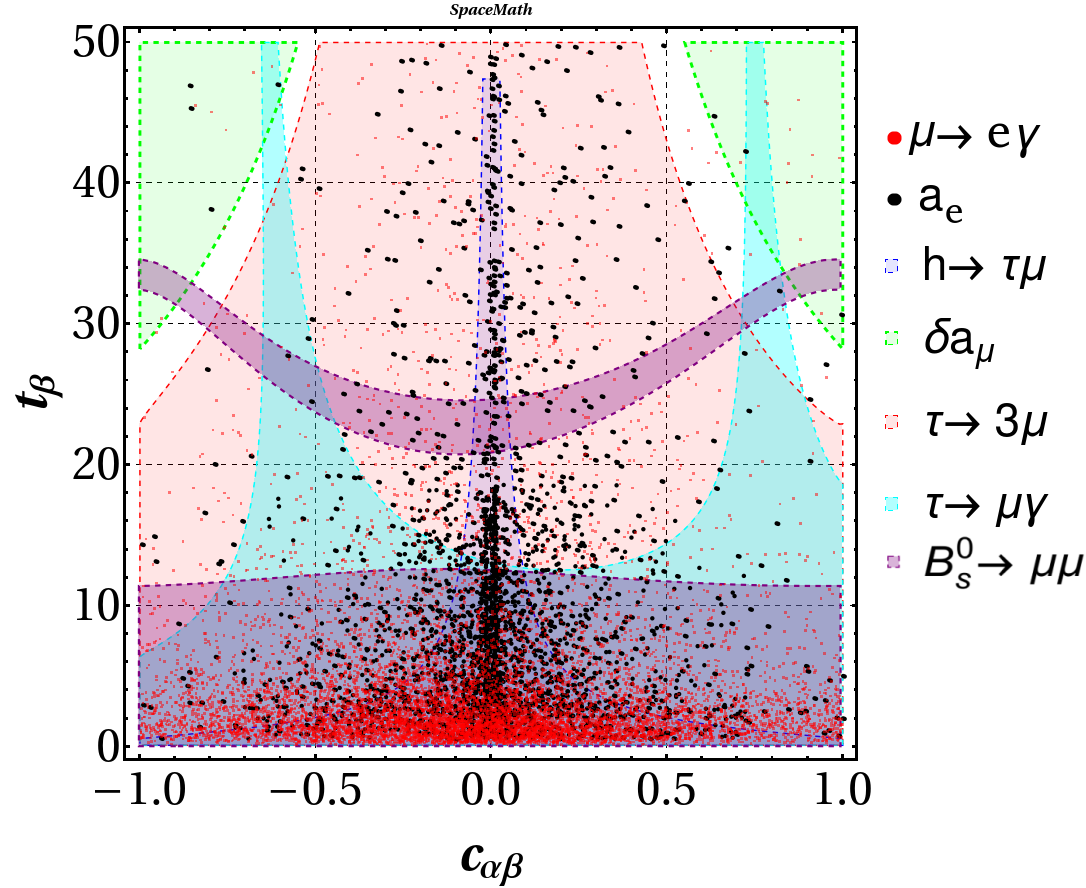}}\label{2b}}
\subfigure[]{{\includegraphics[scale=0.2]{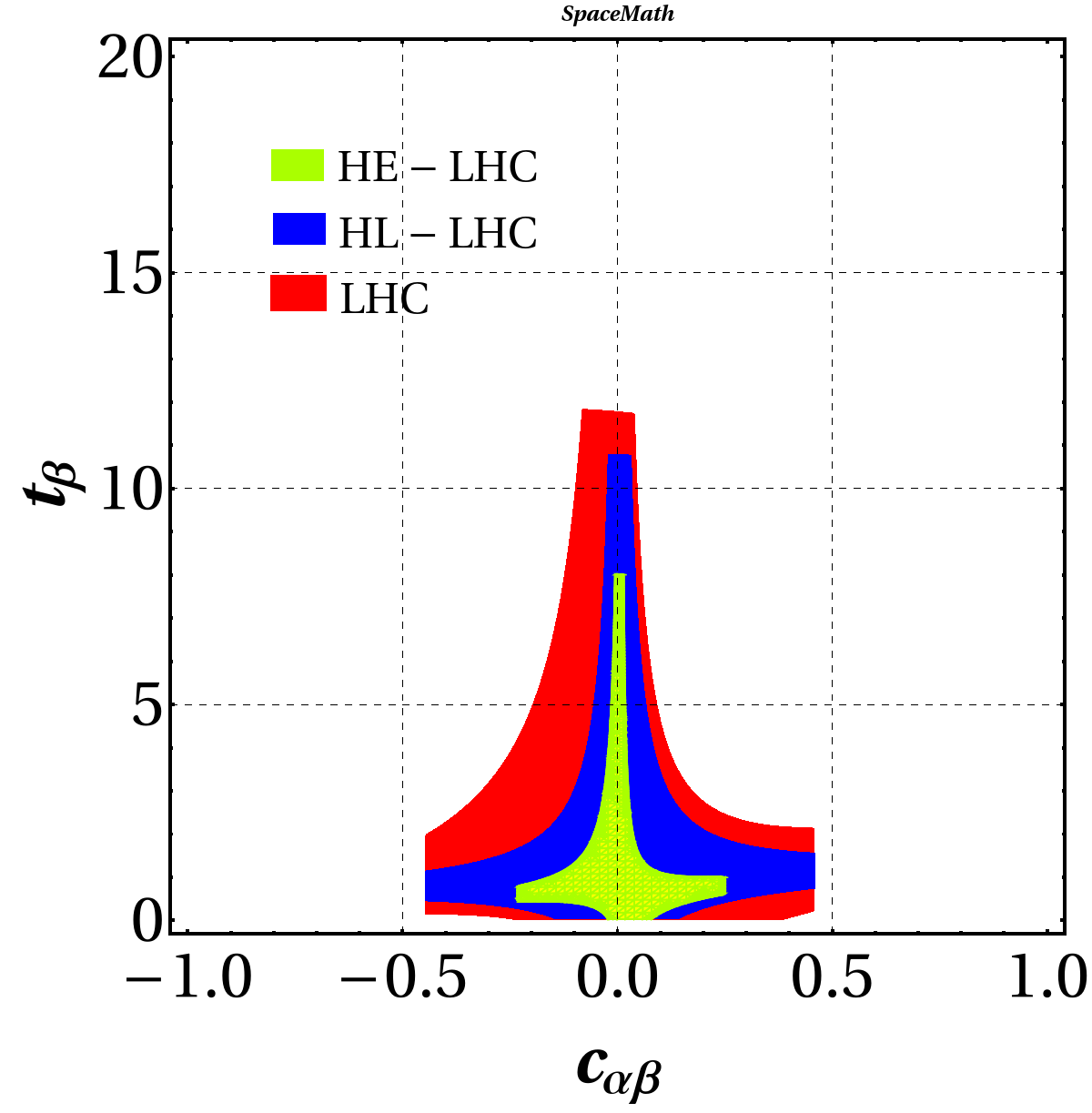}}\label{2c}}
 \caption{The shadowed areas represent the allowed regions in the plane $c_{\alpha\beta}$-$t_{\beta}$: (a) Coupling modifiers $\kappa$-factors, (b) $\mu\to e \gamma$, electric dipole moment of the electron $a_e$, $h\to\tau\mu$, $\delta a_\mu$, $\tau\to 3\mu$, $\tau\to\mu\gamma$ and $B_s^0\to\mu^+\mu^-$, (c) intersection of all allowed regions. We have also included projections for HL-LHC and HE-LHC. }
 \label{tb_cab} 
\end{figure}

In Eq.~\eqref{kappa}, $\Gamma(\phi_i\to X)$ is the decay width of $\phi_i=h^{\text{2HDM-III}}$, $h^{\text{SM}}$ into $X=b\bar{b},\,\tau^-\tau^+,\,ZZ,\,WW,\,\gamma\gamma$, $gg$. Here $h^{\text{2HDM-III}}$ corresponds the SM-like Higgs boson coming from 2HDM-III and $h^{\text{SM}}$ stands for the SM Higgs boson; $\sigma(pp\to \phi_i)$ is the Higgs boson production cross section through proton-proton collisions. Figure~\ref{2a} shows colored areas corresponding to the allowed regions by each channel $X$. As far as the observables $h\to \tau\mu$, $\delta a_{\mu}$, $a_e$,  $\tau\to 3\mu$, $\tau\to\mu\gamma$, $\mu\to e\gamma$ and $B_s^0\to \mu^-\mu^+$ are concerned, we present in Figure~\ref{2b} the area in accordance with experimental bounds\footnote{All the necessary formulas to perform our analysis of the model parameter space are reported in the Appendix of Ref.~\cite{Arroyo-Urena:2018mvl}, the manual of \texttt{SpaceMath}~\cite{Arroyo-Urena:2020qup} and a paper of one of us~\cite{Arroyo-Urena:2017sfb}.}. Meanwhile, Figure~\ref{2c} displays the region consistent with all constrains. We also include the projections for HL-LHC and HE-LHC for Higgs boson data~\cite{Cepeda:2019klc}, and for the decay $B_s^0\to\mu^+\mu^-$~\cite{ATLAS:2019icc}. 

Once we consider all the observables, we find strong restrictions for the 2HDM-III parameter space on the $c_{\alpha\beta}-t_{\beta}$ plane, in particular we observe that $c_{\alpha\beta}= 0.05$ admits values for $t_{\beta}\approx 2$, $t_{\beta}\approx 8$, $t_{\beta}\approx 10$  for HE-LHC, HL-LHC and LHC, respectively. A special case, the alignment limit, {\it i.e.}, $c_{\alpha\beta}\approx 0$ allows $t_{\beta}\approx 12,\,11,\,8$ for the LHC, HL-LHC and HE-LHC respectively. A fact to highlight is that the 2HDM-III is able to accommodate the current discrepancy between the experimental measurement and the theoretical SM prediction of the muon anomalous magnetic dipole moment $\delta a_{\mu}$. However, from Figure~\ref{2b}, we observe that the allowed region by $\delta a_{\mu}$ is out of the intersection of the additional observables. This happens by choosing the parameters shown in Table~\ref{ParValues}. We find that $\delta a_{\mu}$ is sensitive to the parameter changing flavor $\chi_{\tau\mu}$, which was fixed to the unit in order to obtain the best fit of the 2HDM-III parameter space. Under this choice, $\delta a_{\mu}$ is explained with high values of $t_{\beta}$.
%

\subsubsection{Constraint on $m_H$ and $m_A$}\label{constraintmhmasandmhpm}
\begin{figure}[!ht]
\center{\includegraphics[scale=0.7]{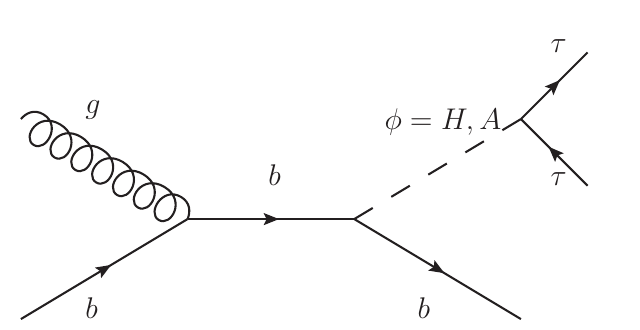}}
 \caption{Feynman diagram of the production of $\phi$ in association with a bottom quark at LHC, with a subsequent decay into $\tau\tau$ pair. }
 \label{FeynmanDiagram2} 
\end{figure}

\begin{figure}[!ht]
\centering
\subfigure[ ]{\includegraphics[scale=0.32,angle=270]{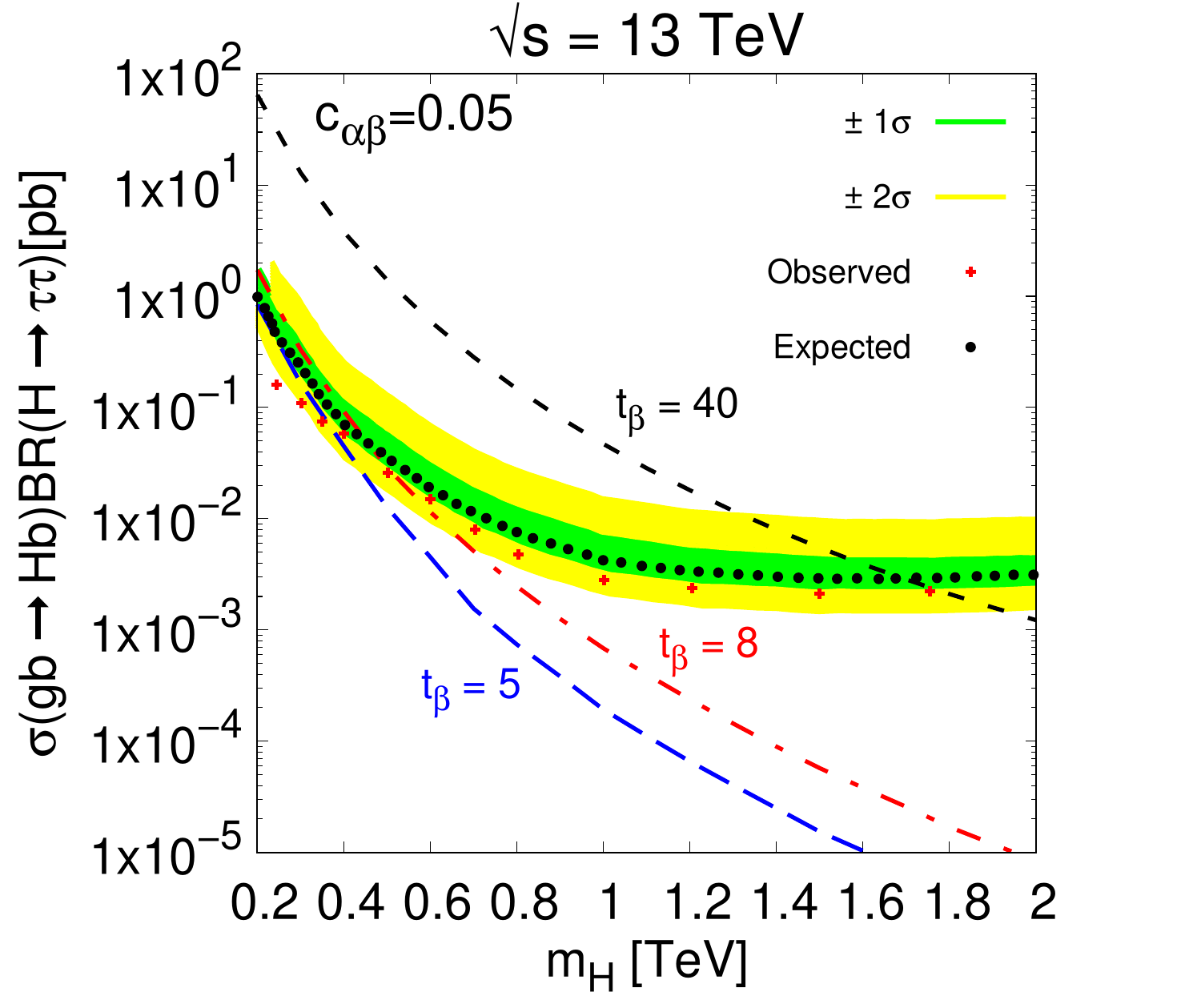}\label{XS_H_timesBRHtautau}}
\subfigure[ ]{\includegraphics[scale=0.32,angle=270]{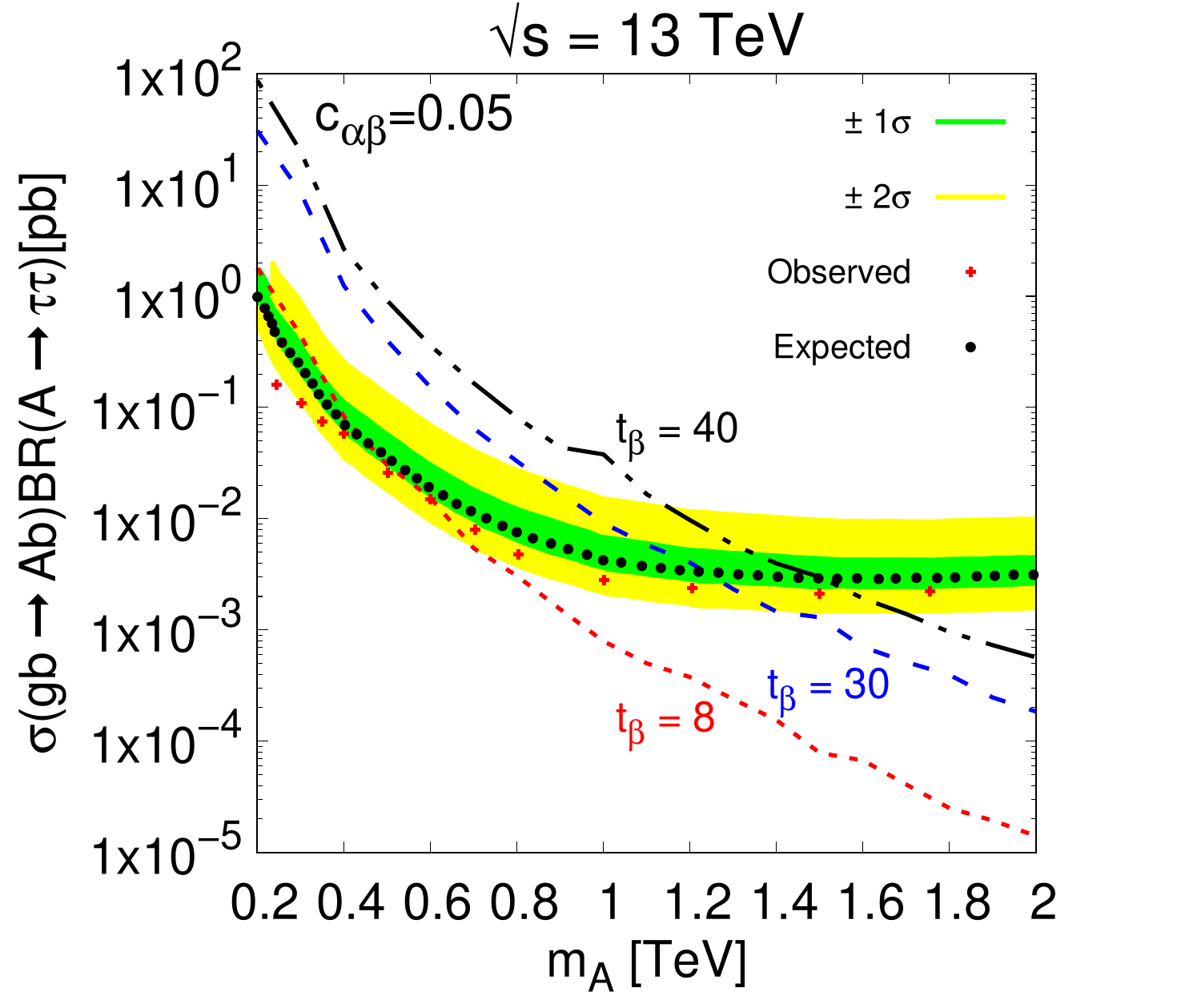}\label{XS_A_timesBRHtautau}}
 \caption{The observed and expected at 95$\%$ CL upper limits on the production cross section times ditau branching ratio for a scalar boson produced via $b$-associated production as a function of (a) the $\mathcal{CP}$-even mass for $t_{\beta}=$5, 8, 40 and (b) the $\mathcal{CP}$-odd mass for $t_{\beta}=$ 8, 30, 40. We take $c_{\alpha\beta}=0.05$. }
 \label{fig:productioncrosssection}
\end{figure}

The ATLAS and CMS collaborations reported results of a search for hypothetical neutral heavy scalar and pseudoscalar bosons in the ditau decay channel~\cite{ATLAS:2017nxi,CMS:2018rmh}. The search was done through the process $gb\to\phi\to\tau\tau$, where $\phi=A,\,H$ (see Figure~\ref{FeynmanDiagram2}). Nevertheless, no significant deviation was observed from the predicted SM background, but upper limits on the production cross section $\sigma(gb\to\phi)$ times branching ratio $\mathcal{BR}(\phi\to\tau\tau)$ were imposed.
Figure~\ref{XS_H_timesBRHtautau} shows the $\sigma(gb\to Hb)\times\mathcal{BR}(H\to\tau\tau)$ as a function of $m_H$ for illustrative values of $t_{\beta}=5,\,8,\,40$ and $c_{\alpha\beta}=0.01$, while in Figure~\ref{XS_A_timesBRHtautau} we present the same process but as a function of $m_A$ for $t_{\beta}=8,\,30,\,40$. The red crosses and black points correspond to the observed and expected values at 95$\%$ C.L., respectively. The green (yellow) band represents the interval at $\pm 1 \sigma$ ($\pm 2 \sigma$) according to the expected value. We implement the Feynman rules in $\texttt{CalcHEP}$~\cite{Belyaev:2012qa} to evaluate such a process. 

From Figure~\ref{XS_H_timesBRHtautau} we can notice that $m_H\lesssim 500$ GeV ($m_H\lesssim 700$ GeV) are excluded at 1$\sigma$ (2$\sigma$) when $t_{\beta}= 8$, while for $t_{\beta}\lesssim 5$ the upper limit on $\sigma(gb\to \phi b)\times\mathcal{BR}(\phi\to\tau\tau)$ is easily accommodated. Despite $t_{\beta}=$40 is ruled out (see Figure~\ref{2c}), we include it to illustrate the consistency (in the high mass regime) with the results reported in previous section. As far as the pseudoscalar mass is concerned, from Figure~\ref{XS_A_timesBRHtautau}, we find that $m_A\lesssim$ 600 GeV ($m_H\lesssim$ 700 GeV) are excluded at 1$\sigma$ (2$\sigma$) for $t_{\beta}= 8$. 

\subsubsection{Constraint on the charged scalar mass $m_{H^\pm}$}\label{subsubsection:constraintoncharguedscalar}
The detection of a charged scalar $H^{\pm}$ would represent a clear signature of new physics. Constraints were obtained from collider searches for the $H^{\pm}$ production and its subsequent decay into a $\tau^{\pm}\nu_\tau$ pair~\cite{ATLAS:2018gfm}. More recently the ATLAS collaboration reported a study on the charged Higgs boson produced either in top-quark decays or in association with a top quark. Subsequently the charged Higgs boson decays via $H^{\pm}\to\tau^{\pm}\nu_{\tau}$ with a center-of-mass energy of 13 TeV~\cite{ATLAS:2018gfm}. However, we find that such processes are not a good way to constrain the charged scalar mass $m_{H^{\pm}}$ predicted in 2HDM-III. Nevertheless, the situation is opposite if one consider the decay $b\to s\gamma$ which imposes severe lower limits on $m_{H^\pm}$ because to the charged boson contribution~\cite{Ciuchini:1997xe, Misiak:2017bgg}.

\begin{figure}[!htb]
\centering{\includegraphics[scale=0.35,angle=270]{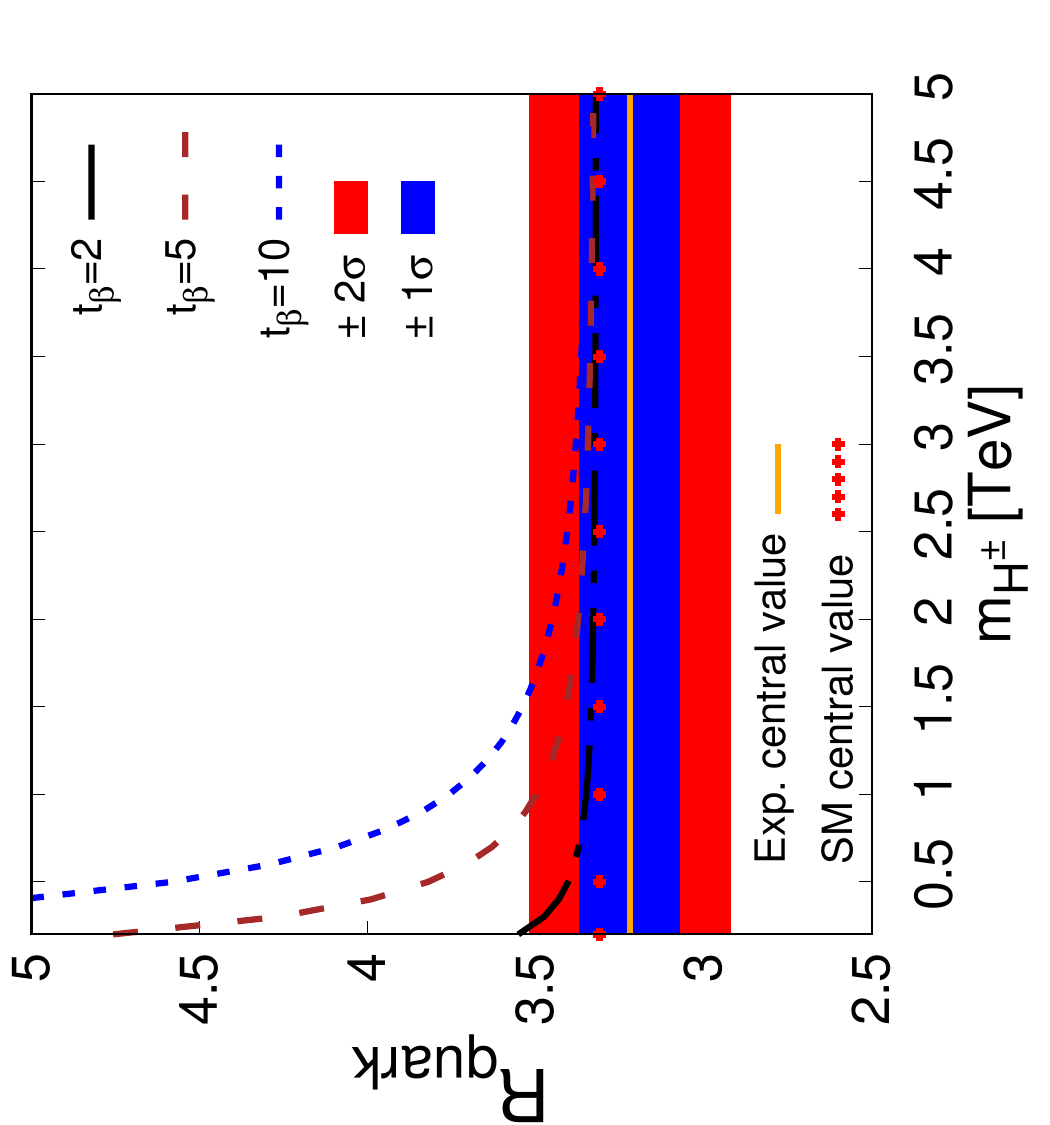}}
 \caption{$R_{quark}$ at NLO in QCD as a function of $m_{H^\pm}$ for $t_{\beta}=2,\, 5,\,  10$. Solid line represents the experimental central value while red area indicate the theoretical SM, around the  central value. Blue and red bands stand for 1$\sigma$ and 2$\sigma$, respectively. $R_{quark}$ is defined in Eq.~\ref{eq:rquark}.}
  \label{mCH}  
\end{figure}

Figure~\ref{mCH} shows the ratio $R_{quark}$ as a function of $m_{H^\pm}$ for $t_{\beta} = 2, 5, 10$. Here, $R_{quark}$ is given by:
\begin{equation}
R_{quark}=\dfrac{\Gamma(b\to X_s\gamma)}{\Gamma(b\to X_ce\nu_e)}.
\label{eq:rquark}
\end{equation} 
When $t_{\beta}=2$, we notice that  $100\,\text{GeV}\lesssim m_{H^{\pm}}$ ($700\,\text{GeV}\lesssim m_{H^{\pm}}$) is ruled out, at $2\sigma$ ($1\sigma$). The intermediate value $t_{\beta}=5$ excludes masses $800\,\text{GeV}\lesssim m_{H^{\pm}}$ ($2\,\text{TeV}\lesssim m_{H^{\pm}}$). Meanwhile, $t_{\beta}=10$ imposes a stringent lower bound $1.6\,\text{TeV}\lesssim m_{H^{\pm}}$ ($3.2\,\text{TeV}\lesssim m_{H^{\pm}}$). 

In summary, we present in Table~\ref{ParValues} the values used in the subsequent calculations.
\begin{table}[!ht]
\caption{Values of the parameters used in the calculations.}
\begin{center}
\begin{tabular}{cc}
\hline 
Parameter & Value\tabularnewline
\hline 
\hline 
$c_{\alpha\beta}$ & 0.01\tabularnewline
\hline 
$t_{\beta}$ & 0.1-8\tabularnewline
\hline 
$\chi_{\tau\mu}$ &1\tabularnewline
\hline 
$m_H=m_A$ & 800 GeV \tabularnewline
\hline 
\end{tabular}
\par\end{center}
\label{ParValues}
\end{table}
%

\subsection{Collider analysis for $h \rightarrow \tau \mu$}\label{subsec:collideranalysis}
\begin{figure}[!ht]
	\centering
	\includegraphics[scale = 0.35, angle=270]{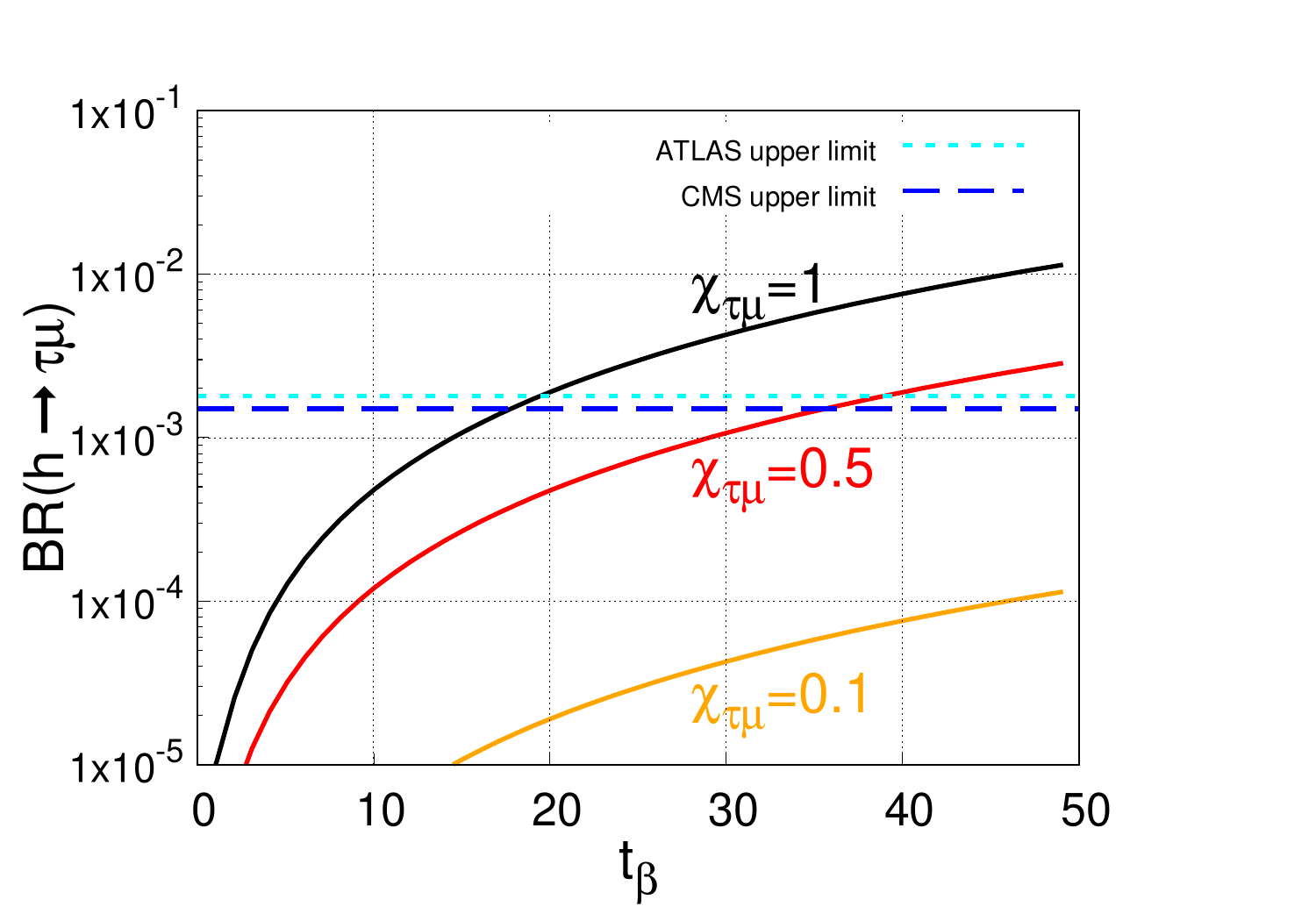}
	\caption{Branching ratio of the decay $h\to\tau\mu$ as a function of $t_{\beta}$ for $\chi_{\tau\mu}=0.1,\,0.5,\,1$. The horizontal line represents the upper limit on $\mathcal{BR}(h\to\tau\mu)$.}
	\label{BRhtaumu}
\end{figure}

In this section we simulate the production of the SM-like Higgs boson $h$ with its subsequent decay into a $\tau\mu$ pair at the LHC, HL-LHC, HE-LHC and FCC-hh.
Let us first show in Figure~\ref{BRhtaumu}, the branching ratio of the $h \to \tau \mu$  decay, $\mathcal{BR}(h\to\tau\mu)$, as a function of $t_{\beta}$ for $\chi_{\tau \mu}=0.1,\,0.5,\,1$ and $c_{\alpha\beta}=0.01$. We also include the upper limit reported by ATLAS and CMS collaborations~\cite{ATLAS:2023mvd, CMS:2021rsq}. We notice that for $\chi_{\tau \mu}=1$, values of $t_\beta \leq 17$ are allowed, while $\chi_{\tau \mu}=0.5$ allows $t_\beta \leq 35$, values of $ t_\beta \leq 50$ are admissible when $\chi_{\tau \mu}=0.1$.

We analyze the LFV Higgs signal for the options of three cases corresponding to Future Hadron Colliders (FHC), namely: HL-LHC, HE-LHC and FCC-hh, with the following luminosities:
\begin{itemize}
		\item \textbf{FHC1}: HL-LHC at a center-of-mass energy of 14 TeV and integrated luminosities 0.3 through 3 ab$^{-1}$,
		\item \textbf{FHC2}: HE-LHC at a center-of-mass energy of 27 TeV and integrated luminosities in the interval 0.3-12 ab$^{-1}$,
		\item \textbf{FHC3}: FCC-hh at a center-of-mass energy of 100 TeV and integrated luminosities from 10 to 30 ab$^{-1}$.
\end{itemize}
%


Once we have under control the free model parameters, as shown previously, we now turn to estimate the number of signal events produced in the three accelerator options. 

\begin{figure}[!ht]
	\centering
	\subfigure[ ]{\includegraphics[scale = 0.21, angle=270]{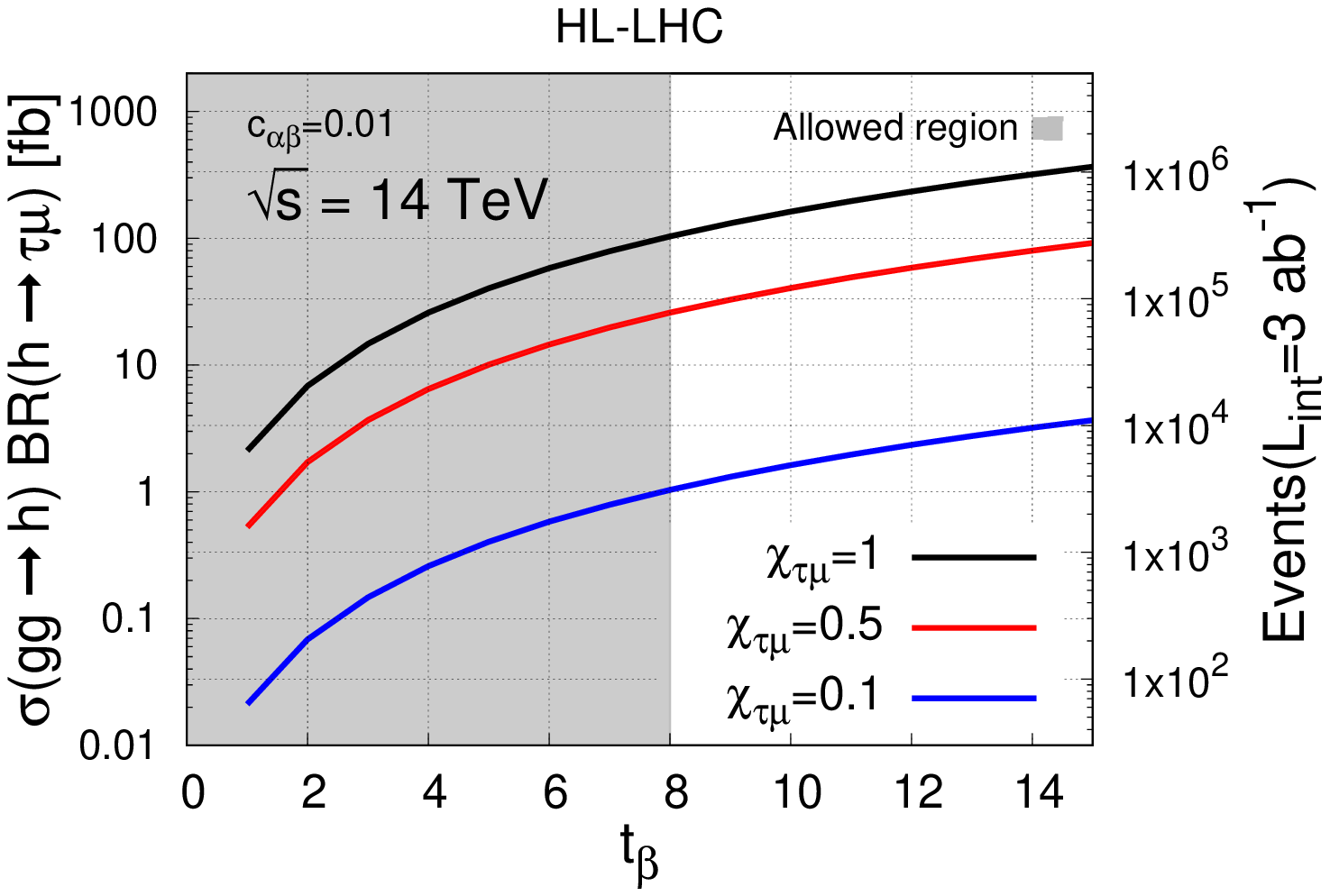}}
	\subfigure[ ]{\includegraphics[scale = 0.21, angle=270]{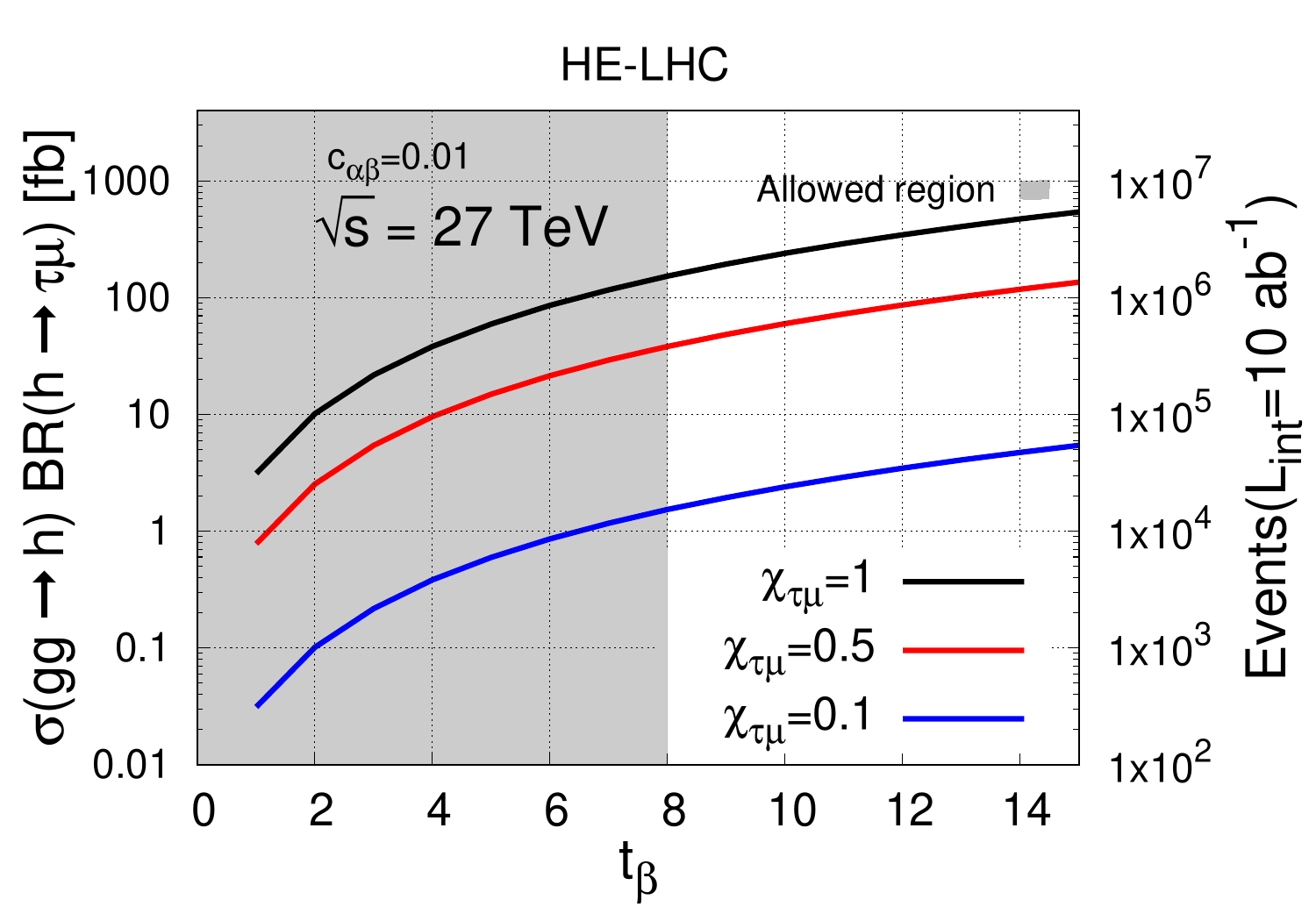}}
	\subfigure[ ]{\includegraphics[scale = 0.21, angle=270]{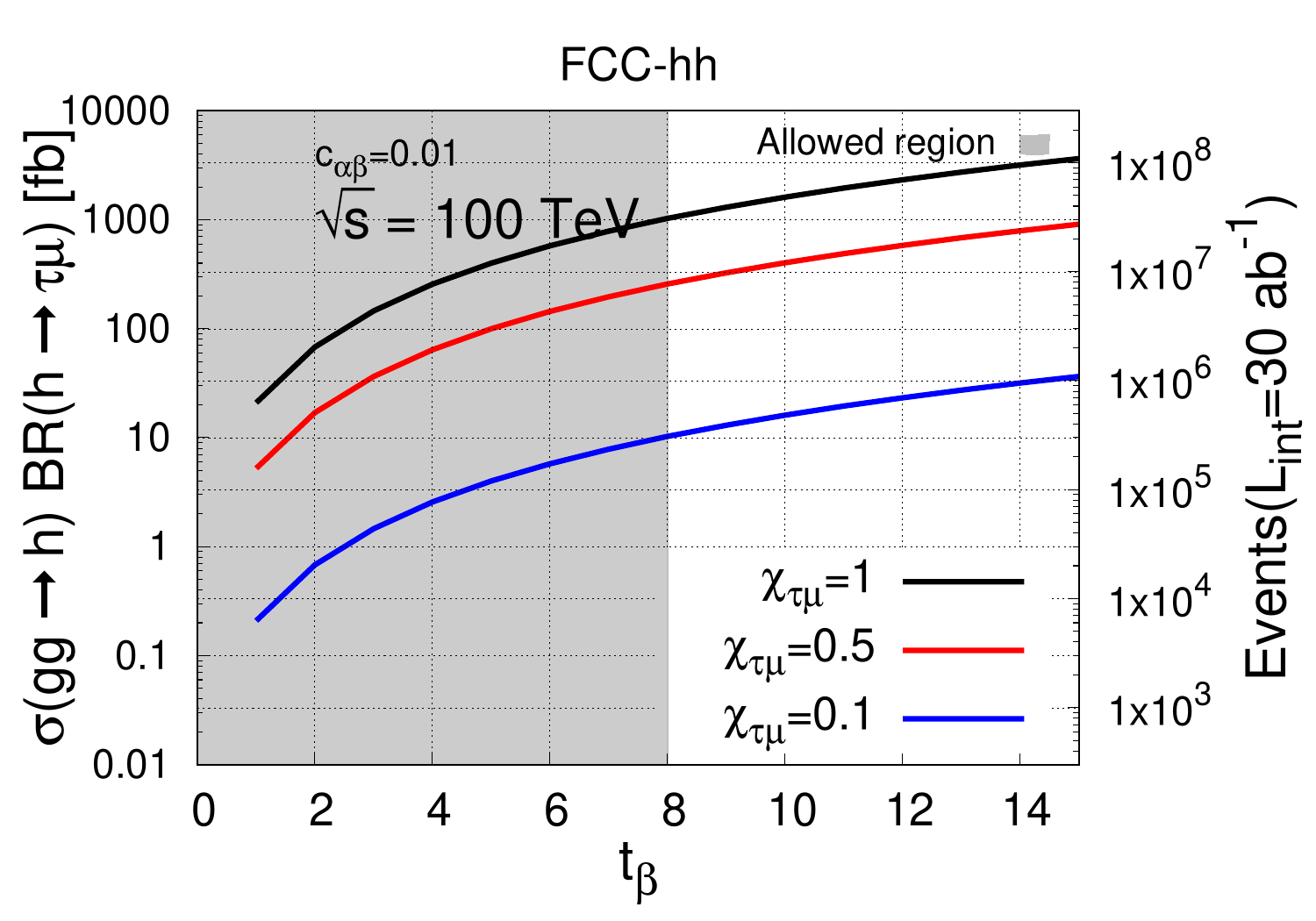}}
	\caption{(a) Scenario \textbf{FHC1}, (b) Scenario \textbf{FHC2}, (c) Scenario \textbf{FHC3} for $c_{\alpha \beta}=0.01$. Left axis: $\sigma(gg\to h)\mathcal{BR}(h\to\tau\mu)$ as a function of $t_{\beta}$ for $\chi_{\tau\mu}$=0.1, 0.5, 1. Right axis: events-$t_{\beta}$ plane. The dark area corresponds to the allowed region. See Table~\ref{ParValues}. }
\label{Events}
\end{figure}

Figure~\ref{Events} shows the product of $\sigma(gg\to h)\mathcal{BR}(h\to\tau\mu)$ as a function of $t_{\beta}$ (left axis) and number of events (right axis) for $\textbf{FHC1}$, $\textbf{FHC2}$ and $\textbf{FHC3}$. The dark area indicates the allowed region by all observables analyzed previously (see Table~\ref{ParValues}). Given the center-of-mass energy and the integrated luminosity, we find that the maximum number of signal events produced, denoted by $\mathcal{N}_{\mathcal{S}}^{\textbf{FHCi}}$, are of the order $\mathcal{N}_{\mathcal{S}}^{\textbf{FHC1}}=\mathcal{O}(10^5)$, $\mathcal{N}_{\mathcal{S}}^{\textbf{FHC2}}=\mathcal{O}(10^6)$, $\mathcal{N}_{\mathcal{S}}^{\textbf{FHC3}}=\mathcal{O}(10^7)$. Here, we consider $t_{\beta}=8$ and $\chi_{\tau\mu}=1$.
%

Let us now to analyze the signature of the decay $h\to\tau\mu$, $\Gamma(h\to\tau\mu)$, with $\tau\mu=\tau^-\mu^++\tau^+\mu^-$ and its potential SM background. The CMS and ATLAS collaborations~\cite{CMS:2015qee, ATLAS:2016joj} searched for this process through two $\tau$ decay channels: electron decay $\tau\to e\nu_{\tau}\nu_{e}$, $\Gamma(\tau\to e\nu_{\tau}\nu_{e})$, and hadron decay $\tau_h\mu$. In this work, we will study the former. As far as our computation methods are concerned, we first implement the model in $\texttt{LanHEP}$~\cite{Semenov:2014rea} for exporting the UFO files necessary to be interpreted in $\texttt{MadGraph5}$~\cite{Alwall:2011uj}, later it is interfaced with \texttt{Pythia8}~\cite{Sjostrand:2008vc} and \texttt{Delphes 3}~\cite{deFavereau:2013fsa} to simulate the response of the detector. Subsequently, we generate signal and background events, the last ones at NLO in QCD. We used $\texttt{NNPDF}$ parton distribution functions~\cite{NNPDF:2017mvq}.

\subsection*{Signal and SM background processes}
The signal and background processes are as following:
\begin{itemize}
		\item \textit{SIGNAL-} We search for a final state coming from the process $gg\to h\to\tau\mu\to e\nu_{\tau}\nu_{e}\mu$. This channel contains exactly two charged leptons, namely, one electron (or positron) and one antimuon (or muon) and missing energy because of undetected neutrinos.
		\item \textit{BACKGROUND-} The potential SM background for the signal come from: 
		\begin{enumerate}
			\item Drell-Yan process, followed by the decay $Z\to\tau\tau\to e\nu_{\tau}\nu_{e}\mu\nu_{\tau}\nu_{\mu}$.
			\item $WW$ production with subsequent decays $W\to e\nu_{e}$ and $W\to \mu\nu_{\mu}$.
			\item $ZZ$ production, later decaying into $Z\to\tau\tau\to e\nu_{\tau}\nu_{e}\mu\nu_{\tau}\nu_{\mu}$ and $Z\to\nu\nu$.
		\end{enumerate}
\end{itemize}
%

\subsection*{Signal significance}
The most restrictive kinematic cut to separate the signal from the background processes is the transverse mass $M_T$, defined as the invariant mass of the vector sum, the third component being zero:
%
%
%
	\begin{equation}\label{MT}
		M_T^{\mu e}\equiv M_T=\sqrt{\Big(\sqrt{a_x^2+a_y^2}+\sqrt{b_x^2+b_y^2}\Big)^2-\Big(a_x+b_x\Big)^2-\Big(a_y+b_y\Big)^2}.
\end{equation}

\begin{figure}
		\centering
		\subfigure[ ]{\includegraphics[scale=0.28]{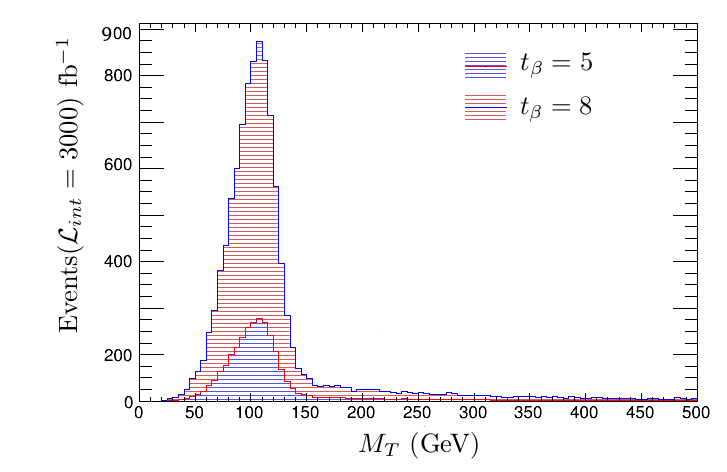}}
		\subfigure[ ]{\includegraphics[scale=0.28]{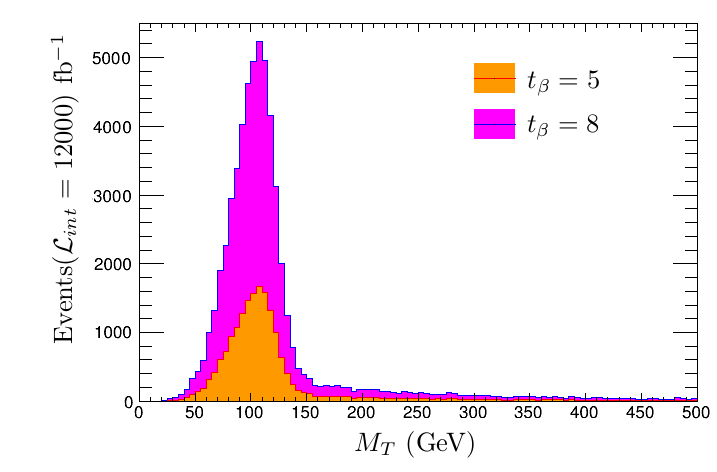}}
		\subfigure[ ]{\includegraphics[scale=0.28]{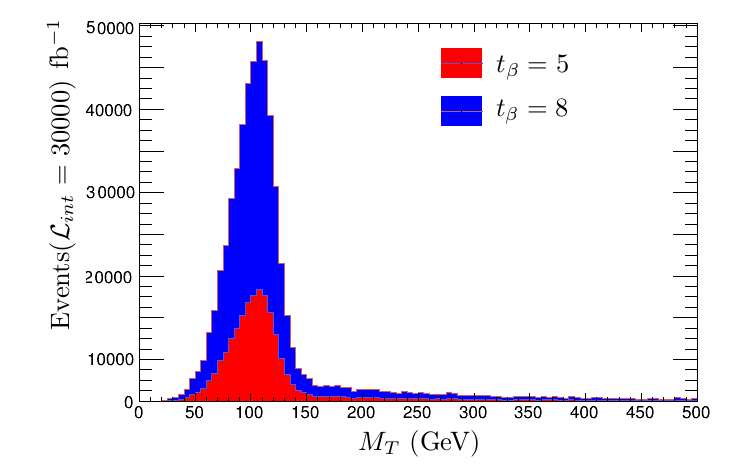}}
		\caption{Distribution of the transverse mass $M_T$ (Eq.~\eqref{MT}) for scenarios: (a) $\textbf{FHC1}$, (b) $\textbf{FHC2}$ and (c) $\textbf{FHC3}$. \label{MassCOLINEAR}}
	\end{figure}

\begin{figure}[!htb]
	\centering
	\subfigure[]{\includegraphics[scale = 0.23]{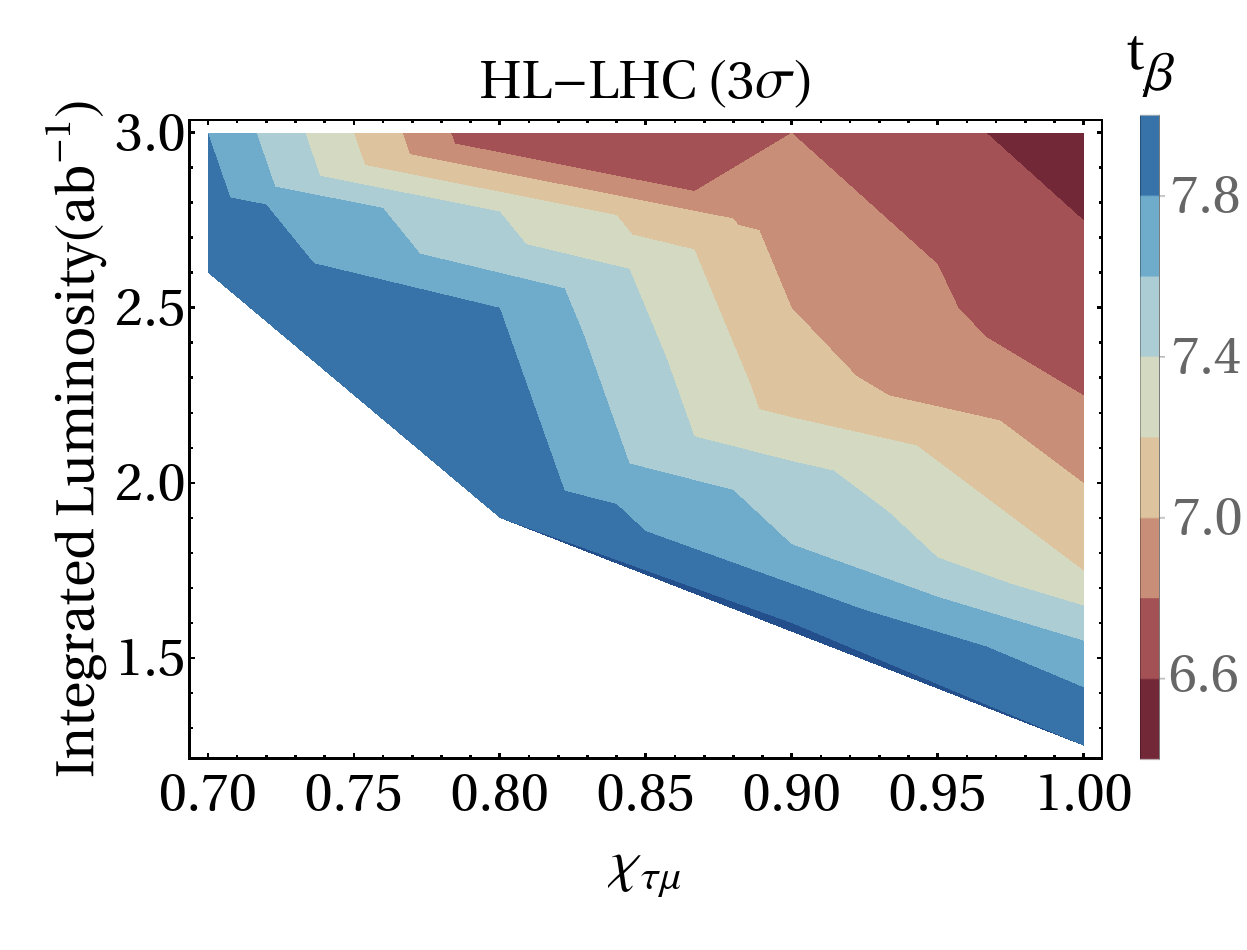}}\\
	\subfigure[]{\includegraphics[scale = 0.23]{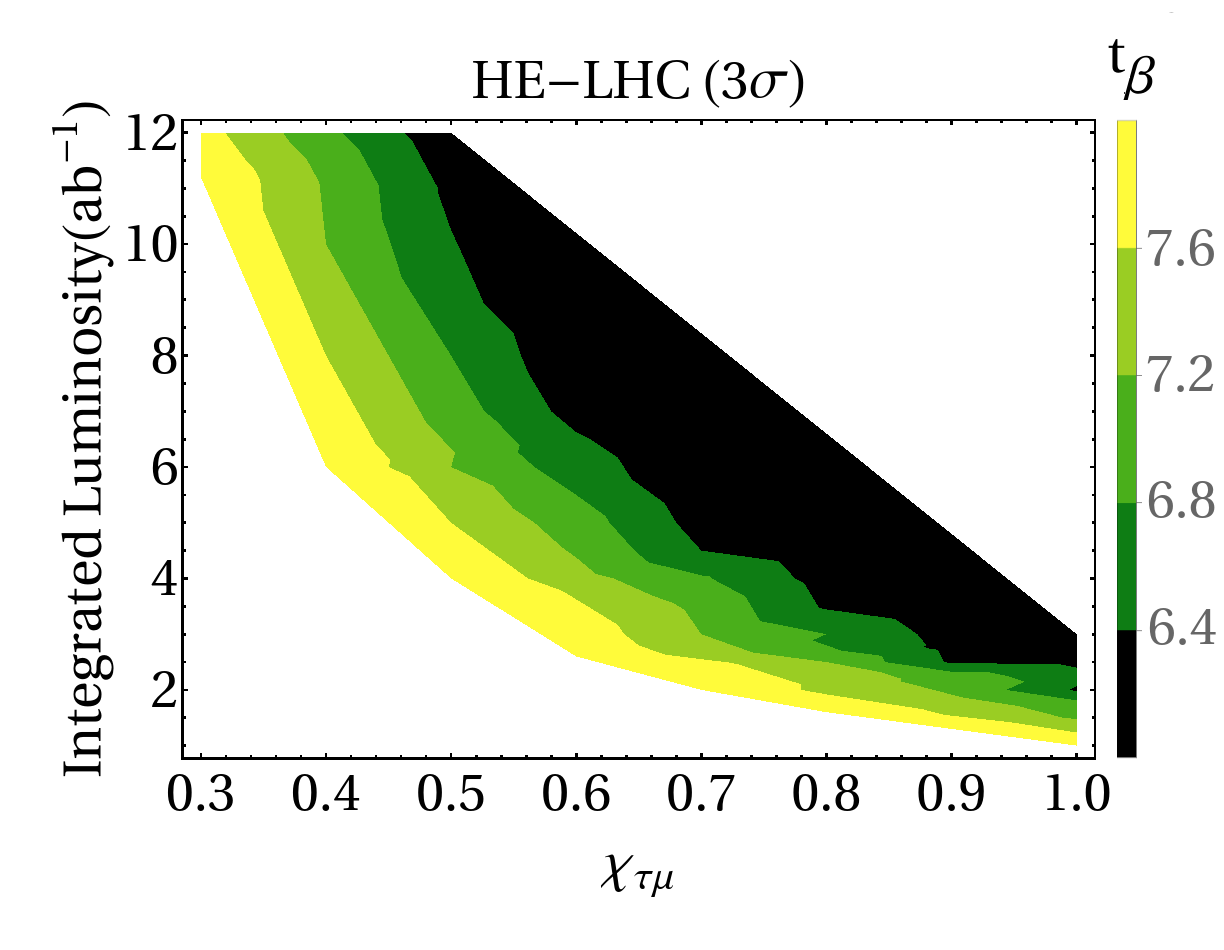}}
	\subfigure[]{\includegraphics[scale = 0.23]{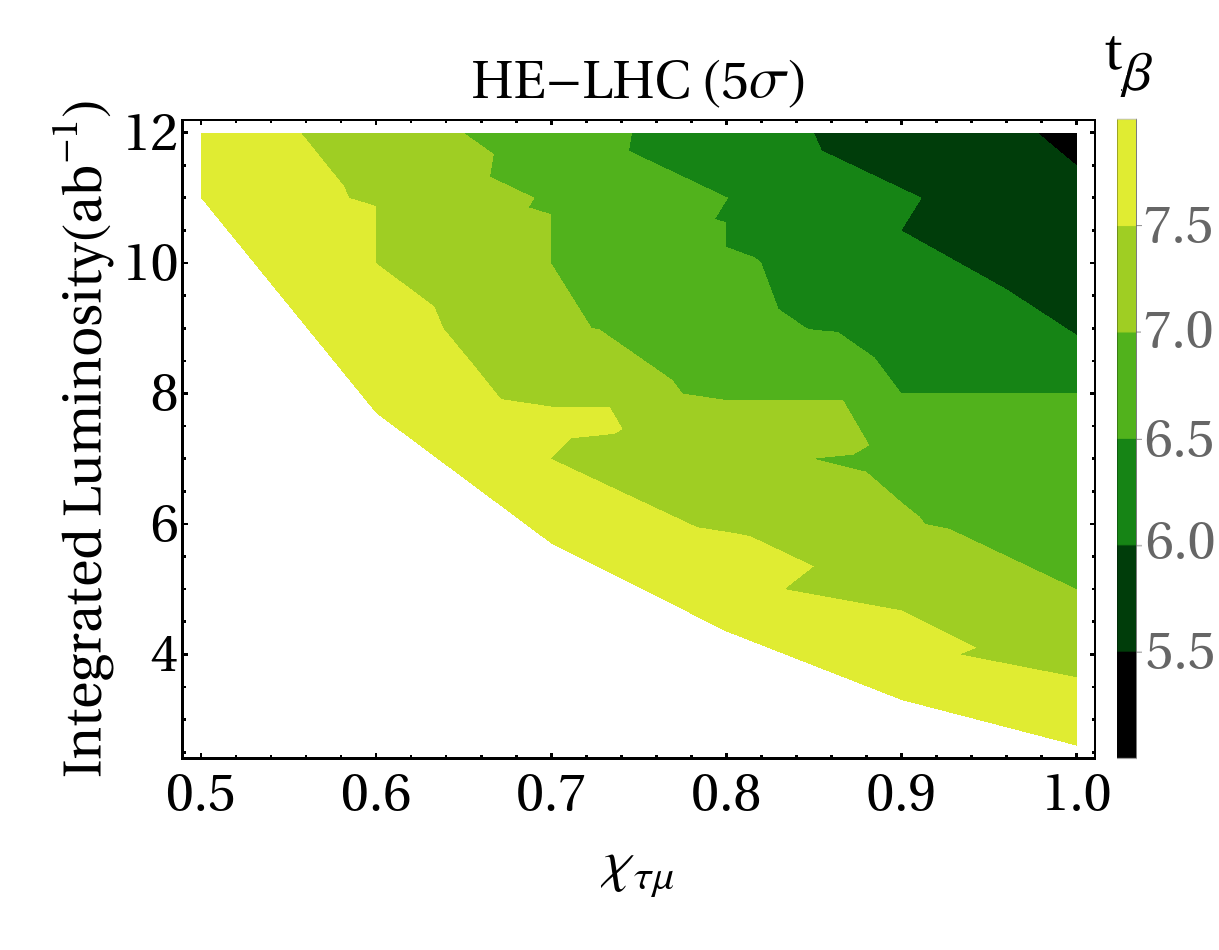}}
	\subfigure[]{\includegraphics[scale = 0.23]{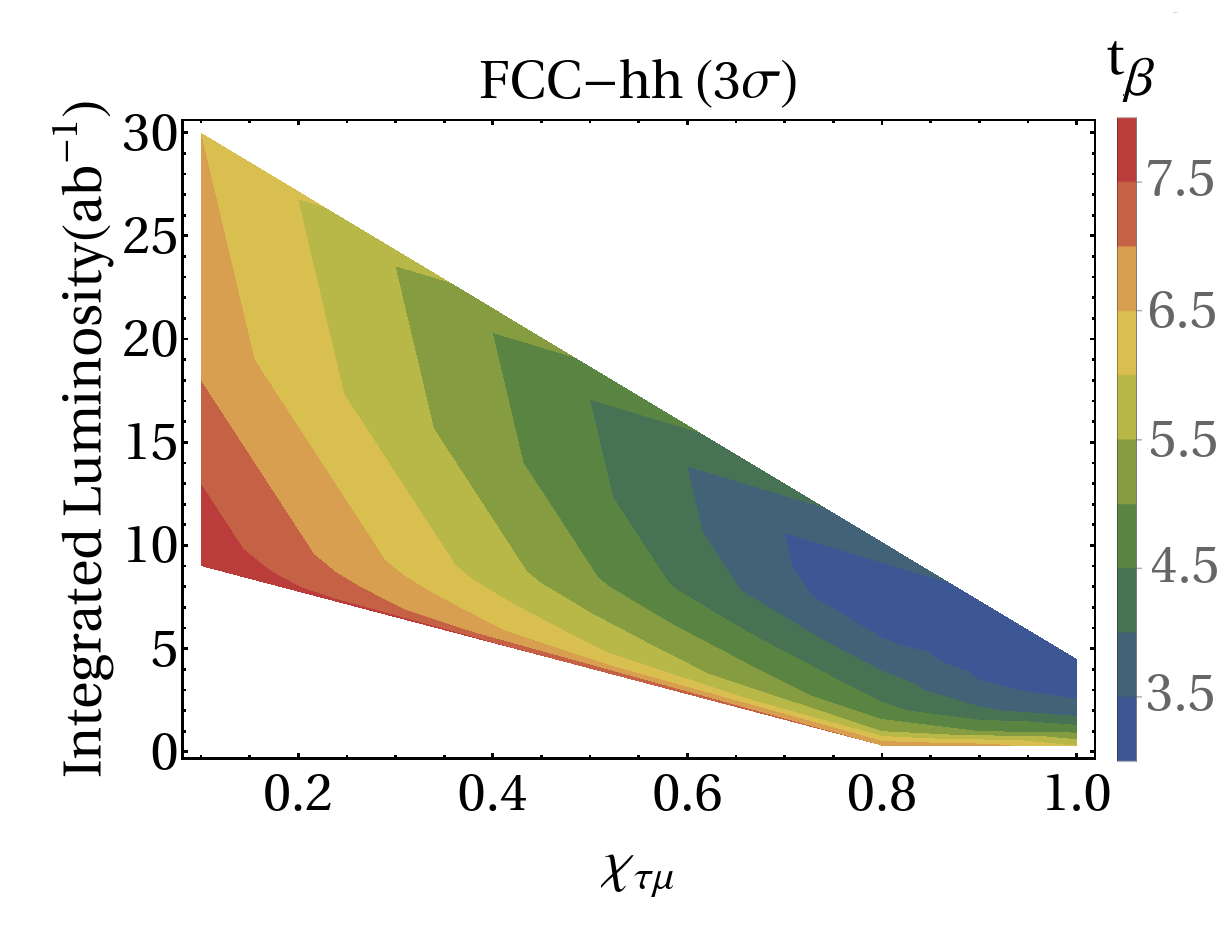}}
	\subfigure[]{\includegraphics[scale = 0.23]{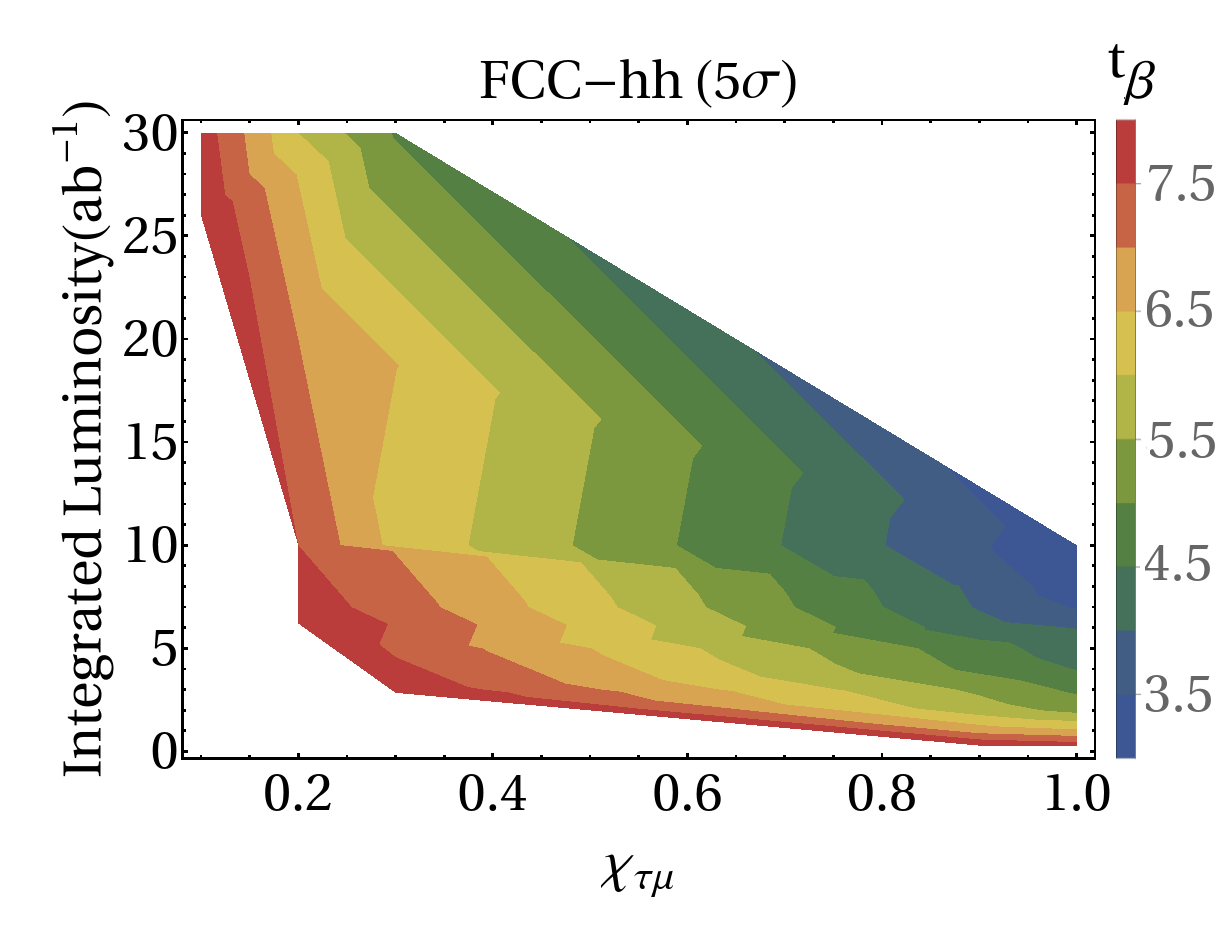}}
	\caption{Signal significance as a function of the integrated luminosity, the $\chi_{\tau\mu}$ parameter of LFV coupling $h \tau \mu$ and $t_\beta$. (a), (b), (d): 3$\sigma$ and (c), (e): 5$\sigma$.  \label{ALLscenarios}}
\end{figure}

We present in Figure~\ref{MassCOLINEAR} the distribution of collinear mass of the signal events for the scenarios (a) $\textbf{FHC1}$, (b) $\textbf{FHC2}$ and (c) $\textbf{FHC3}$, We consider integrated luminosities of 3, 12 and 30 ab$^{-1}$ respectively, and $t_{\beta}$= 5, 8.
	For the kinematic analysis of the processes, we use the package $\texttt{MadAnalysis5}$~\cite{Conte:2012fm}. 	
	Besides the collinear mass and transverse mass, we applied additional cuts for both signal and background~\cite{ATLAS:2016joj, CMS:2015qee} as shown in Table~\ref{KinematicCutsSA} (scenario \textbf{FHC1}). Scenarios \textbf{FHC1-FHC3} are available electronically in~\cite{Kopp:2016rzs}. The number of the signal ($\mathcal{N}_S$) and background ($\mathcal{N}_B$) after the kinematic cuts were applied, are also included. We consider to the signal significance defined as $\mathcal{N}_S/\sqrt{\mathcal{N}_S+\mathcal{N}_B}$; once we compute it, the efficiency of the kinematic cuts for both signal and background are: $\epsilon_S\approx 0.13$ and $\epsilon_B\approx 0.014$, respectively.
\begin{table}
	\caption{Kinematic cuts applied to the
		signal and main SM background for scenario $\mathbf{FHC1}$.}\label{KinematicCutsSA}
	\begin{center}
		\begin{tabular}{|c|c|c|c|c|}
			\hline 
			Cut number & Cut & $\mathcal{N}_S$ & $\mathcal{N}_B$ & $\mathcal{N}_S/\sqrt{\mathcal{N}_S+\mathcal{N}_B}$
			\tabularnewline
			\hline 
			\hline 
			& Initial (no cuts) & $5.77\times 10^4$ & $2.00\times 10^8$ & $4.10$\tabularnewline
			\hline 
			$1$ & $|\eta^{e}|<2.3$ & $2.53\times 10^4$ & $1.30\times 10^8$ & $2.20$\tabularnewline
			\hline 
			$2$ & $|\eta^{\mu}|<2.1$ & $1.64\times 10^4$ & $1.07\times 10^8$ & $1.59$\tabularnewline
			\hline 
			$3$ & $0.1<\Delta R(e,\,\mu)$ & $1.63\times 10^4$ & $1.06\times 10^8$ & $1.59$\tabularnewline
			\hline 
			$4$ & $10<p_{T}(e)$ & $1.55\times 10^4$ & $3.90\times 10^7$ & $2.49$\tabularnewline
			\hline 
			$5$ & $20<p_{T}(\mu)$ & $1.21\times 10^4$ & $2.04\times 10^7$ & $2.69$\tabularnewline
			\hline 
			$6$ & $10<\text{MET}$  & $1.12\times 10^4$ & $2.01\times 10^7$ & $2.50$\tabularnewline
			\hline 
			$8$ & $110<M_{T}(e \mu)<140$ & $8.67\times 10^3$ & $4.83\times 10^6$ & $4.83$\tabularnewline
			\hline 
			\end{tabular}
		\par\end{center}
\end{table}
In the previous analysis, we found that the LHC presents difficulties in carrying out an experimental scrutiny, achieving a statistical significance of $\sim 1.4\sigma$. However, more promising expectations arise at future stages of the LHC, namely, HL-LHC in which a prediction about $4.6\sigma$ for 3 ab$^{-1}$ and $t_{\beta}=8$ is found. Meanwhile, the HE-LHC (FCC-hh) offers exceptional results, we find a signal significance around $5.04\sigma$ ($\sim 5.43\sigma$) considering 9 ab$^{-1}$ and $t_{\beta}=6$ (15 ab$^{-1}$ and $t_{\beta}=3$). Then, we present in Figure~\ref{ALLscenarios} regions for the model parameters corresponding to a potential evidence ($3\sigma$) or a possible discovery ($5\sigma$) for the three scenarios \textbf{FHC1}, \textbf{FHC2}, \textbf{FHC3}.

%
%
%


\section{One-loop calculation of $h  \to \ell_a \ell_b$ within 2HDM-I,II, with massive neutrinos}\label{section:oneloop}

In the context of the 2HDM type I and II, the interaction $h \ell_a \ell_b$ with $a \neq b$ is not allowed at tree level, because of all fermions interact with only one Higgs doublet, such as in the SM. However, considering neutrino masses the LFV-HD are induced at one loop by means of charged currents. Here, we consider the Inverse SeeSaw (ISS) mechanism to describe the lightness of neutrino masses~\cite{PhysRevLett.56.561, PhysRevD.34.1642, BERNABEU1987303}, which is a natural extension of the type I and II variants of the 2HDM to include neutrino masses. The ISS mechanism allows to describe neutrino masses with a low scale $\mathcal{O}$(TeV) heavy neutrinos reachable at current colliders. In contrast, the SeeSaw type I (SS) mechanism describe lightness of neutrino masses incorporating heavy neutrinos with mass scale about $10^{14}$ GeV~\cite{PhysRevLett.44.912, PhysRevD.22.2227}, which are currently not accessible by experiments. The main difference of ISS in comparison to SS is the new scale $\mu_X$ which comes from the new extra singlets, in this case, the lightness of neutrinos is a consequence a low mass scale for $\mu_{X}$ and a moderately large $M_R$ mass scale (see Eq.~\eqref{ec:seesaw_2HDM}). Heavy neutrinos in the ISS have been linked in the past to non-negligible LFV higgs decays caused by radiative corrections~\cite{PhysRevD.91.015001, THAO2017159, PhysRevD.95.095029}.
In both versions of the 2HDM that we take into consideration here, the Yukawa term proportional to $\Phi_2$ is taken into account by the Dirac mass term for neutrinos. All fermions interact with $\Phi_2$ in type I, and up-type fermions (in this case, neutrinos) interact with $\Phi_2$ in type II.

In this work, we assume a normal hierarchy for light neutrino masses and we collect the experimental values for the neutrino oscillations parameters in the Table~\ref{tb:oscilaciones}~\cite{Esteban:2018azc}.
{\renewcommand{\arraystretch}{1.4}
	\begin{table}
		\begin{center}
			\begin{tabular}{ccc}
				\toprule
				& BFP $\pm 1 \sigma$   & $3 \sigma$ \\ 
				\hline
				$\sin^2{(\theta_{12})}$& $0.310^{+0.013}_{-0.012}$ & $0.275-0.350$  \\ 
				\hline
				$\sin^2{(\theta_{23})}$& $0.582^{+0.015}_{-0.019}$  & $0.428 - 0.624$  \\ 
				\hline
				$\sin^2{(\theta_{13})}$& $0.02240^{+0.00065}_{-0.00066}$ & $0.02244 - 0.02437$ \\ 
				\hline
				$\dfrac{\Delta m_{21}^2}{10^{-5} \text{eV}^2}$& $7.39  ^{+0.021}_{-0.020}$ & $6.79 - 8.01$  \\ 
				\hline
				$\dfrac{\Delta m_{31}^2}{10^{-5} \text{eV}^2}$&$2.525  ^{+0.033}_{-0.031}$  & $2.431 - 2.622$ \\ 
				\hline
			\end{tabular}
		\end{center}
		\caption{Neutrino oscillations data for a normal hierarchy. The second column show the Best Fit Point (BFP) and the $1\sigma$ range, third column shows the  $3\sigma$ range.}
		\label{tb:oscilaciones}
\end{table}}
%
In the normal hierarchy, we can rewrite the masses of $\nu_{2,3}$ in terms of $m_1$, the mass of the lightest neutrino $\nu_1$,  and the square mass differences as follows
\begin{equation}\label{ec:light_mnu}
m_i = \sqrt{m_1^2 + \Delta m_{i1}^2 },\quad i=2,3.
\end{equation}
%

\subsection{The model $\nu$2HDM-I, II}\label{subsec:themodel}
%
%

In the Yukawa sector, we include three pairs of fermionic singlets ($\nu_R$, $X$) to the SM~\cite{PhysRevD.91.015001}. The right handed neutrinos $\nu_R$ singlets partners of $\nu_L$ in the SM with lepton number $L = -1$, while the extra singlet $X$ has lepton number $L = 1$. With these additional fermions, we can write the Lagrangian as
\begin{equation}
\mathcal{L}^\nu = - Y_{ij}^{2 ,\nu} \overline{L_i} \tilde{\Phi}_2 \nu_{Rj} - M_{ij}^\nu \overline{\nu_{Ri}^C}X_j - \dfrac{1}{2}\mu_{Xij}\overline{X_{i}^C}X_j + \textrm{H. c.},
\end{equation}
where $i,j= 1,2,3$, $Y^{\nu}_2$ is the $3\times 3$ neutrino Yukawa matrix associated to the $\Phi_2$ Higgs doublet, $M_R$ is a lepton number conserving complex $3\times 3$ matrix and $\mu_X$ is a Majorana complex $3\times 3$ symmetric mass matrix that violates the lepton number  conservation by two units. Hence, the $9\times 9$ neutrino mass matrix after the electroweak symmetry breaking, in the $(\nu_L, \nu_{Ri}^C, X_i^C)$ basis, is given by
\begin{equation}
M^\nu_{\text{ISS}} = 
\begin{pmatrix}
0 & m_{D}^2 & 0 \\
m_{D}^{2,\top} & 0 & M_R \\
0 & M_R^\top &\mu_X
\end{pmatrix},
\end{equation}
where $m_{D}^2 = \dfrac{v_2}{\sqrt{2}}Y^{2, \nu}$ is the Dirac matrix, which is symmetric then can be diagonalized by a $9\times 9$ unitary matrix $U_\nu$ given by 
\begin{equation}
U_\nu^T M_{\mathrm{ISS}}^\nu U_\nu=\operatorname{diag}\left(m_{n_1}, \ldots, m_{n_9}\right),
\end{equation}
where $m_{n_i}$ are the masses of the nine physical Majorana neutrinos ${n_i}$ and are associated with the electroweak basis by the rotation
\begin{equation}
\left(\begin{array}{c}
\nu_L^C \\
\nu_R \\
X
\end{array}\right)=U_\nu P_R\left(\begin{array}{c}
n_1 \\
\vdots \\
n_9
\end{array}\right), \quad\left(\begin{array}{c}
\nu_L \\
\nu_R^C \\
X^C
\end{array}\right)=U_\nu^* P_L\left(\begin{array}{c}
n_1 \\
\vdots \\
n_9
\end{array}\right).
\end{equation}

As a consequence of neutrino masses, following the notation of Aoki et al.~\cite{AOKI2009} and L. Diaz~\cite{LorenzoDiaz-Cruz:2019imm}, the Yukawa Lagrangian in the leptonic sector is  given by
\begin{align}\notag
\mathcal{L}^{2HDM}_{\text{Yuk,ISS}} &= -\sum_{\ell}\left(\dfrac{m_{\ell}}{v}c_h^\ell \overline{\ell}\ell h + \dfrac{m_{\ell}}{v}c_H^\ell \overline{\ell}\ell H - i \dfrac{m_{\ell}}{v}e_A^\ell \overline{\ell}\gamma_5\ell A \right) \\  \notag
&\quad - \dfrac{1}{v}\sum_{ij}^9 \left(c^n_h \overline{n_i}(\omega_{ij}P_R + \omega_{ij}^* P_L)n_j h + 
c^n_H \overline{n_i}(\omega_{ij}P_R + \omega_{ij}^* P_L)n_j H  - i e^n_A \overline{n_i}(\omega_{ij}P_R - \omega_{ij}^* P_L)n_j A\right) \\  \notag
&\quad - \left( 
\dfrac{\sqrt{2}U_{n \ell}}{v}\overline{n}(m_n e^n_A P_L + m_\ell e^\ell_A P_R)\ell H^+  + \text{H.c.}\right) \\ \label{eq:LY2HDM}
&\quad - \left( 
\dfrac{\sqrt{2}U_{n \ell}}{v}\overline{n}(m_n P_L + m_\ell  P_R)\ell G^+  + \text{H.c.}\right) ,
\end{align}
where $P_{L/R}$ are projections operators for left-/right-handed fermions. The second row of Eq.~\eqref{eq:LY2HDM} was calculated in analogy with~\cite{THAO2017159} for the $\nu$SM model. Also, $\omega_{ij} = m_{n_j}C_{ij} + m_{n_i}C_{ij}^*$ and $C_{ij} = \sum_{c=1}^3 U_{ci}^\nu U_{cj}^{\nu *}$. Finally, the $c_{\phi^0}^{f}$ ($\phi^0 = h,H$) and $e_A^{f}$ factors for charged leptons and neutrinos are given in Table~\ref{tab:ce_factors}.
%
%
\begin{table}
	\centering
	\begin{tabular}{ccccc}
		\hline \hline
		\rule[-1ex]{0pt}{2.5ex}   & {\bf Type I} & {\bf Type II} \\
		\hline \hline
		\rule[-1ex]{0pt}{2.5ex} $c_h^\ell$ & $\cos{\alpha}/\sin{\beta}$ & $-\sin{\alpha}/\cos{\beta}$\\
		\hline
		\rule[-1ex]{0pt}{2.5ex} $c_H^\ell$ &
		$\sin{\alpha}/\sin{\beta}$ &
		$\cos{\alpha}/\cos{\beta}$  \\
		\hline
		\rule[-1ex]{0pt}{2.5ex} $e_A^\ell$ &
		$-\cot{\beta}$ & 
		$t_{\beta}$ \\
		\hline
		\rule[-1ex]{0pt}{2.5ex} $c_h^n$ &
		$\cos{\alpha}/\sin{\beta}$ & $\cos{\alpha}/\sin{\beta}$ \\
		\hline
		\rule[-1ex]{0pt}{2.5ex} $c_H^n$ & $\sin{\alpha}/\sin{\beta}$ & $\sin{\alpha}/\sin{\beta}$ \\
		\hline
		\rule[-1ex]{0pt}{2.5ex} $e_A^n$ &
		$\cot{\beta}$ & 
		$\cot{\beta}$ \\ 
		\hline \hline
	\end{tabular}
	\caption{Factors $c_{\phi^0}^{f}$ and $e_A^{f}$ in the $\nu$2HDM-I,II.}
	\label{tab:ce_factors}
\end{table}
The neutral currents in the Lagrangian~\eqref{eq:LY2HDM}, can be rewritten following the notation of Eq.~\eqref{ec:hll_SP}. For charged leptons, only the following $\mathcal{CP}$-conserving and $\mathcal{CP}$-violating factors are non-zero,
%
\begin{equation}
S^{\ell,\phi^0} = \dfrac{g m_\ell}{2 m_W}c_{\phi^0}^{\ell}; \qquad
P^{\ell,A} = -\dfrac{g m_\ell}{2 m_W}e_{A}^{\ell}.
\end{equation}
%

In the case of neutrinos, as a consequence of the ISS, $\mathcal{CP}$-conserving and $\mathcal{CP}$-violating factors appear,
\begin{equation}
S_{ij}^{n,\phi} = \dfrac{1}{2v}c^n_\phi(\omega_{ij} + \omega_{ij}^{*}); \qquad
P_{ij}^{n,\phi} = -\dfrac{i}{2v}c^n_\phi(\omega_{ij} - \omega_{ij}^{*}),
\end{equation}
\begin{equation}
S_{ij}^{n,A} = - \dfrac{1}{2v}e^n_A(\omega_{ij} - \omega_{ij}^{*}); \qquad
P_{ij}^{n,A} = \dfrac{i}{2v}e^n_A(\omega_{ij} + \omega_{ij}^{*}).
\end{equation}
As a result, the interaction of neutral scalars and leptons is given by
%
\begin{align}\label{eq:Lphiff}
\mathcal{L}_{\phi ff} &= 
-\sum_{\ell}\left(S^{\ell, h} \overline{\ell}\ell h + S^{\ell, H} \overline{\ell}\ell H + i P^{\ell, A} \overline{\ell}\gamma_5\ell A \right) \\ 
& - \sum_{ij}^9 \left(\overline{n_i}(S_{ij}^{n,h} + i \gamma_5 P_{ij}^{n,h})n_j h + 
\overline{n_i}(S_{ij}^{n,H} + i \gamma_5 P_{ij}^{n,H})n_j H  + i \overline{n_i}(S_{ij}^{n,A} + i \gamma_5 P_{ij}^{n,A})n_j A\right).
\end{align}

\begin{figure}[!ht]
\centering
{\includegraphics[scale = 0.35]{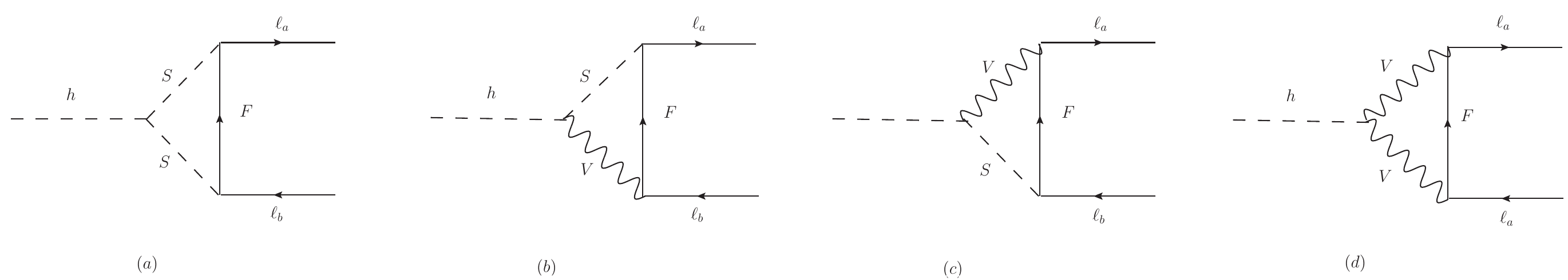}\label{dia1}}
{\includegraphics[scale = 0.4]{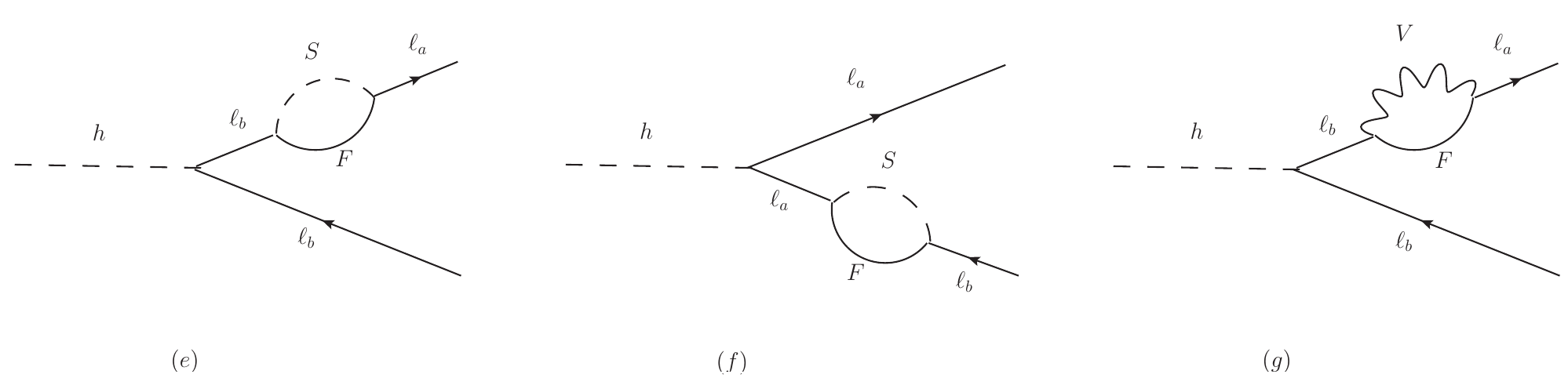}\label{dia2}}
{\includegraphics[scale = 0.4]{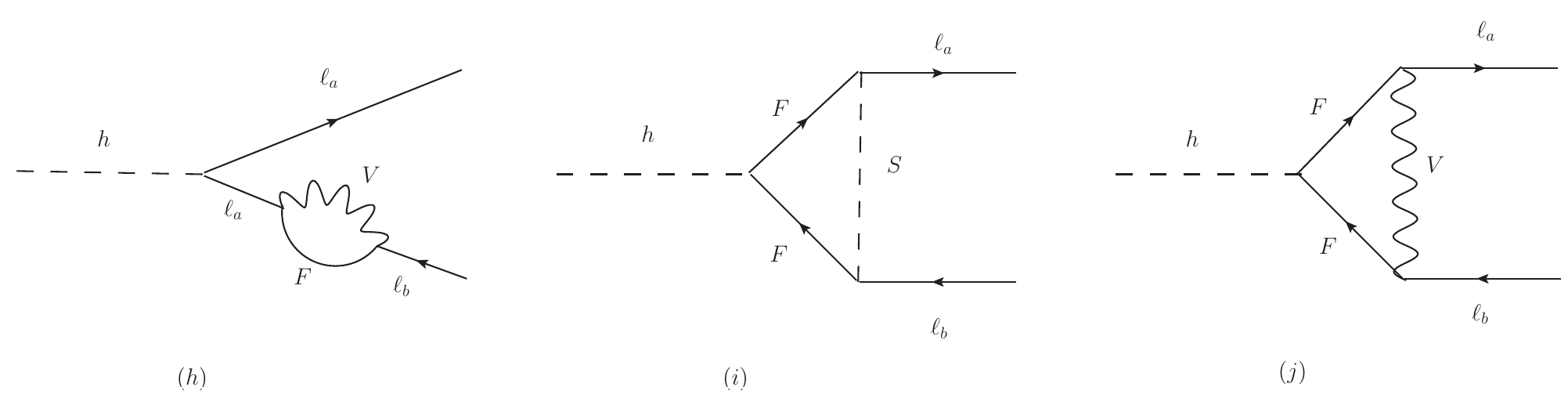}\label{dia3}}
\caption{Feynman diagrams that contribute at one-loop level to the $h\to \ell_a \ell_b$ decay in the 2HDM-I and II~\cite{Zeleny-Mora:2021tym}. Here $F, S, V$ represent fermion, scalar and vector contributions, respectively.}
\label{fig:FeynmanDiagrams}
\end{figure}

The LFV Higgs decays (LFV-HD) are allowed at one-loop by means of charged currents mediated by $G^{\pm}$, $W^{\pm}$ and $H^{\pm}$.
%
Detailed formulae for LFV-HD at one loop was presented in our previous paper~\cite{Zeleny-Mora:2021tym}, with the decay width being written in terms of the form factors $A_{L,R}$, namely:
\begin{align}\label{ec:width}
\notag
\Gamma(h \rightarrow \ell_a \ell_b) &\equiv \Gamma(h \rightarrow \ell_a^{+} \ell_b^{-}) + \Gamma(h \rightarrow \ell_a^{-} \ell_b^{+}) \\ \notag
& = \dfrac{1}{8 \pi m_h} \left[1-\left( \dfrac{m_a^2 + m_b^2}{m_r} \right)^2  \right]^{1/2} \left[1-\left( \dfrac{m_a^2 - m_b^2}{m_r} \right)^2  \right]^{1/2} \\
&\times \left[ (m_h^2 -m_a^2 - m_b^2)(|A_L|^2 + |A_R|^2) - 4 m_a m_b \text{Re}(A_L A^{*}_R) \right],
\end{align} 
with $a,\, b=\mu,\, e, \, \tau \, (a \neq b)$. The general topologies of Feynman graphs that contribute to the amplitude are shown in Figure~\ref{fig:FeynmanDiagrams}, where $S$, $F$ and $V$ denote scalars, fermions and vectors, respectively. In the special case of the $\nu$2HDM-I, II, 18 Feynman graphs contributes to LFV-HD at one loop,  summarized in Table~\ref{tab:2HDM_contributions}, and the associated couplings are given in Table~\ref{tab:couplings_2HDM_seesaw}. We work in Feynman-t'Hooft gauge and use dimensional regularization to manage the divergences. The analytical expressions for the form factors of each diagram are presented in the Appendix~\ref{appen:FormFactors2HDM}. The result for the form factors is finite, {\it i.e.}, free of divergences, as it should be considering that the LFV  Higgs couplings do not exist at tree-level. 
%
%
\subsection{Numerical analysis and results}\label{subsec:numericalanalysis}
The ISS mechanism explain the smallness of neutrino masses, assuming  the hierarchy, $|\mu_X| \ll |M_D| \ll |M_R|$, by means of  
\begin{equation}\label{ec:seesaw_2HDM}
\widehat{m}_\nu \approx M_{D}^2 M^{\top -1}_R \mu_X M^{-1}_R M_{D}^{2,\top},
\end{equation}
where $\mathcal{M} =  M^{\top -1}_R \mu_X M^{-1}_R$. On one hand, the mass matrix for light neutrinos~\eqref{ec:seesaw_2HDM} is proportional to $\mu_X$ and inversely proportional to $M_R$. On the other hand, Eq.~\eqref{ec:seesaw_2HDM} has the usual form in the See-Saw (SS) type I mechanism. Then, considering the flavor basis, where the Yukawa matrix for charged leptons, $Y^\ell$, the mass matrices $M_R$, $\mu_X$, and the gauge interactions are diagonal, as a consequence, the Dirac mass matrix $M_{D}^2$ can be rewritten in terms of 9 neutrino masses, the leptonic mixing matrix $U_{\text{PMNS}}$ and an $3 \times 3$ orthogonal matrix  $Q$ as follows~\cite{CASAS2001171}:
\begin{equation}\label{ec:ISS_MD}
M_{D}^2=\left(\widehat{\mathcal{M}}\right)^{1 / 2} Q \left(\widehat{m}_{\nu}\right)^{1 / 2} U_{\mathrm{PMNS}}^{\dagger},
\end{equation}
where $\widehat{m}_\nu = \text{diag}(m_{n_1},m_{n_2},m_{n_3})$ and $\widehat{\mathcal{M}} = \text{diag}(m_{n_4}, \dots,m_{n_9})$. Although, in general the matrix $Q$ depends on three complex angles. In this work, we chose $Q$ as the identity matrix, also, we assume that $M_R = \text{diag}(M_{R1}, M_{R2}, M_{R3})$ and $\mu_X = \text{diag}(\mu_{X1}, \mu_{X2}, \mu_{X3})$. As a result, $M_{D}^2$ depends only on neutrino masses and the mixing matrix $U_{\text{PMNS}}$. The last one could be calculated using the BFP values from Table~\ref{tb:oscilaciones}. The 9 neutrino masses are reduced to 6 by means of Eq.~\eqref{ec:light_mnu} and assuming $m_{n_1} = 10^{-12}$ GeV. To simplify the numerical analysis we consider degenerate values to  $M_{R1} = M_{R2} = M_{R3} = M_R$ and $\mu_{X1} = \mu_{X2} = \mu_{X3} = \mu_X$.

\subsubsection{Constraint on the THDM-I and THDM-II}\label{subsection:constraintthdmIyII}
We now turn to analyze the THDM-I and THDM-II parameter space
\begin{itemize}
	\item {\bf Signal strength modifiers:}
	We consider the use of signal strength modifiers $\mathcal{R}_X$ to constraint the parameter space. For a production process $\sigma(pp\to\phi_{i})$ and the decay $\phi_i\to X$, $\mathcal{R}_X$ is defined as follows
	\begin{equation}
	\mathcal{R}_X=\dfrac{\sigma(pp\to h) \mathcal{BR}(h\to X)}{\sigma(pp\to h^{\text{SM}}) \mathcal{BR}(h^{\text{SM}}\to X)},
	\end{equation}
	with $\phi_i=h,\,h^{\text{SM}}$, where $h$ stands  for the SM-like Higgs boson coming from an extension of the SM and $h^{\text{SM}}$ is the SM Higgs boson; $\mathcal{BR}(\phi_i\to X)$ is the rate of the $\phi_i\to X$ decay, with 
	$X=b\bar{b},\,\tau^-\tau^+,\,\mu^-\mu^+,\,W^+W^-,\,ZZ,\,\gamma\gamma$. Figure~\ref{ParamSpaceI-II} shows the $c_{\beta \alpha} - t_\beta$ plane for the case of Figure~\ref{Rx_THDM-I} THDM-I and Figure~\ref{Rx_THDM-II} THDM-II ($c_{\beta \alpha} \equiv \cos(\beta - \alpha)$). 
	\begin{figure}[!ht]
		\centering
		\subfigure[ ]{\includegraphics[scale = .3]{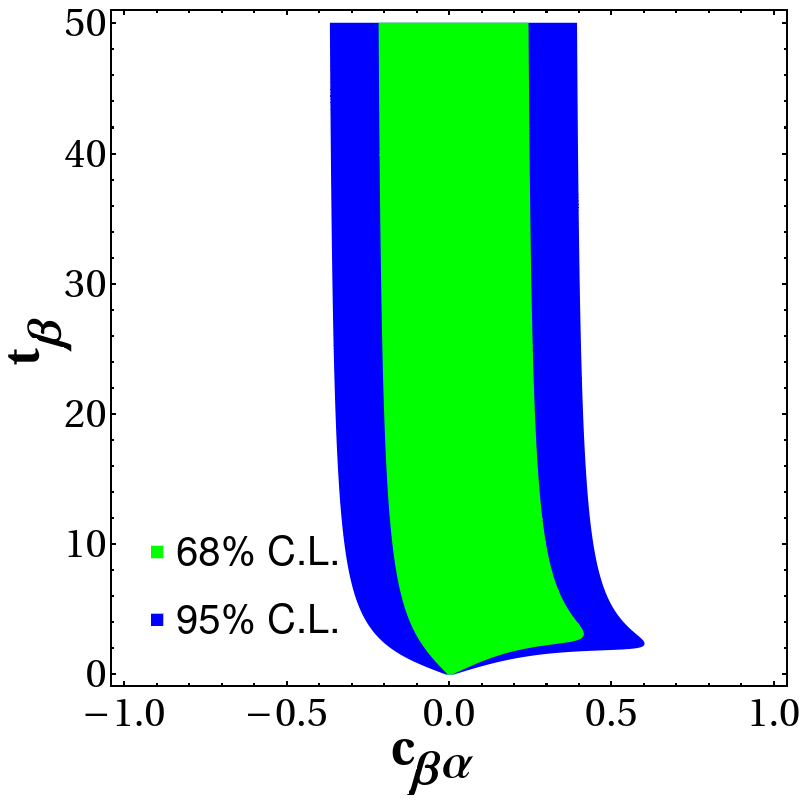}\label{Rx_THDM-I}}
		\subfigure[ ]{\includegraphics[scale = .3]{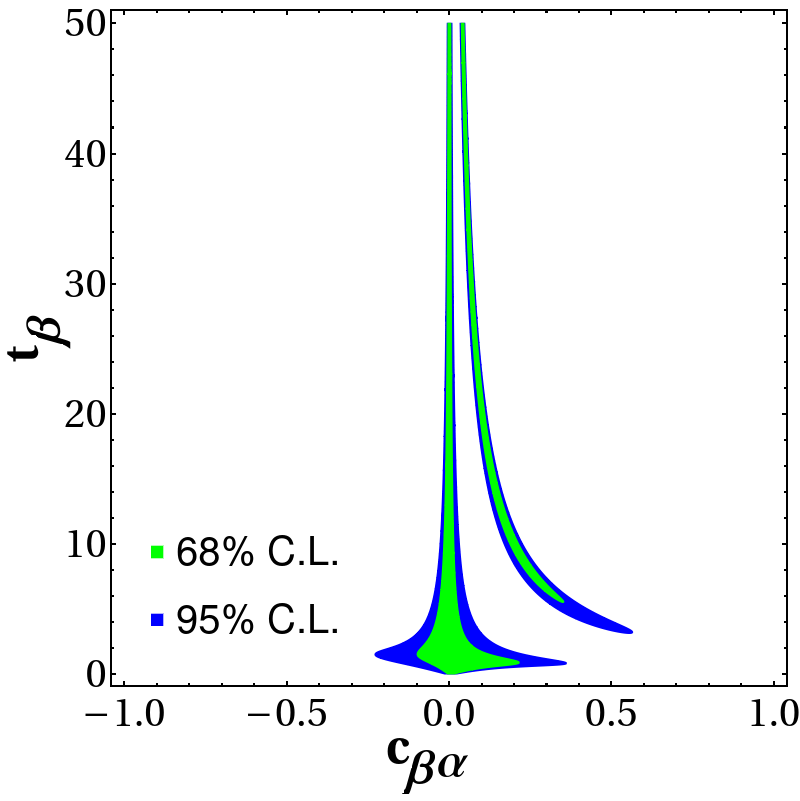}\label{Rx_THDM-II}}
		\caption{The $c_{\beta\alpha}-t_\beta$ shows the allowed areas by LHC Higgs boson data. Green (Blue) area stands for 68\% (95\%) of confidence level.}
		\label{ParamSpaceI-II}
	\end{figure}
	\item {\bf Perturvativity:}
	This bound applies over Yukawa matrix and implies that $|Y^{2, \nu}_{ij}|^2 < 6 \pi$~\cite{PhysRevD.91.015001,THAO2017159}. If we include the factor $(v_2 /\sqrt{2})^2$, we obtain a bound for the Dirac matrix, given by:
	\begin{equation}\label{ec:cota_perturvativity}
	|(M_{D}^2)_{ij}|^2 < 3 \pi v_2^2 = 3 \pi v^2 \sin(\beta)^2.
	\end{equation}
	In the framework that we considered~\eqref{ec:ISS_MD} depends on $M_R$, $\mu_X$ and $\tan{\beta}$, then, the perturbativity bound~\eqref{ec:cota_perturvativity} constraints these parameters as shown in the Figure~\ref{fig:perturvativityboundtb2}.
	\begin{figure}
		\centering
		\includegraphics[width=0.47\linewidth]{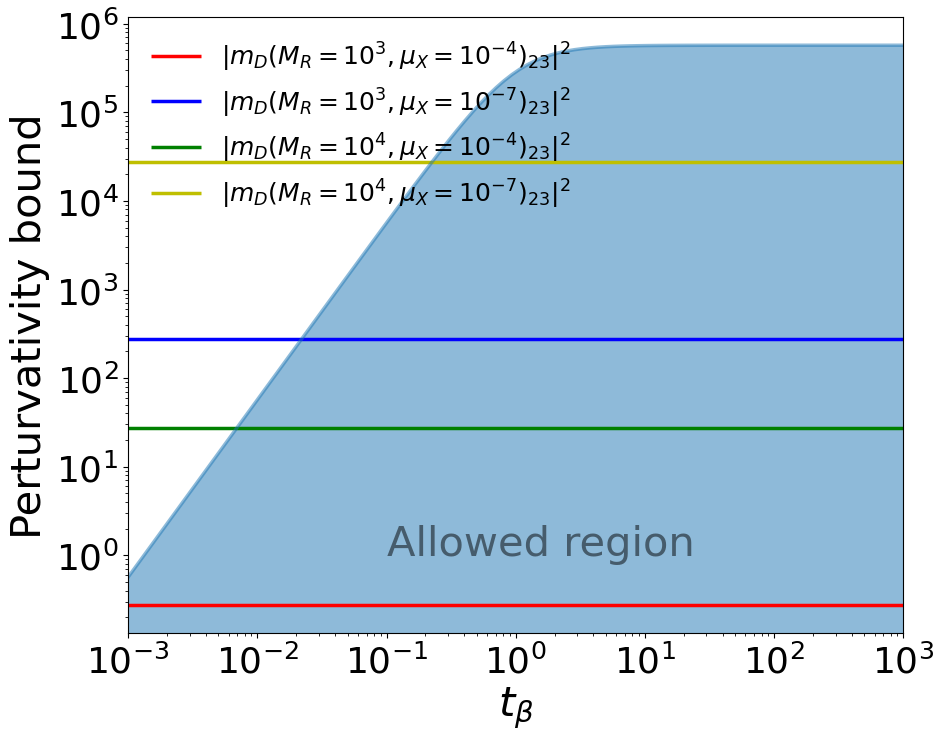}
		\caption{Perturbativity bound on the $(M_D^2)_{23}$ in the $\nu$2HDM-I,II. The blue region represents the allowed perturbativity space, see Eq.~\eqref{ec:cota_perturvativity}.}
		\label{fig:perturvativityboundtb2}
	\end{figure}
\end{itemize}
%

\subsection{Events for the $h \to \mu \tau$ decay at future hadron colliders}\label{subsec:eventsfor}
For the numerical evaluation, we observe that $\mathcal{BR}(h \to \ell_{a} \ell_{b})$ depends on the scales of neutrino masses $M_R$ and $\mu_X$, the parameters $t_{\beta}$ and $c_{\beta \alpha}$, the masses $m_{H^\pm}$, $m_A$ and $\lambda_5$. From the signal strength modifiers, we can observe that the parameter space is in agreement with the alignment limit $\beta - \alpha = \frac{\phi}{2}$ or $c_{\beta  \alpha} = 0$. However, we assume in this work a quasi-alignment value of $c_{\beta \alpha} = 0.01$. Also, we use, $M_R = 1$ TeV, $m_A = 587$ GeV, $\lambda_5 = 2.167$. These values were widely motivated by a deep analysis of the model parameter space through experimental data and theoretical constraints, namely, LHC Higgs boson data, perturbativity and unitarity, respectively. In this sense, we used the \texttt{SpaceMath} package which has implemented machine learning algorithms which predicted the values for $m_A$ and $\lambda_5$.

\begin{figure}[!ht]
\centering
\subfigure[ ]{\includegraphics[scale = .3]{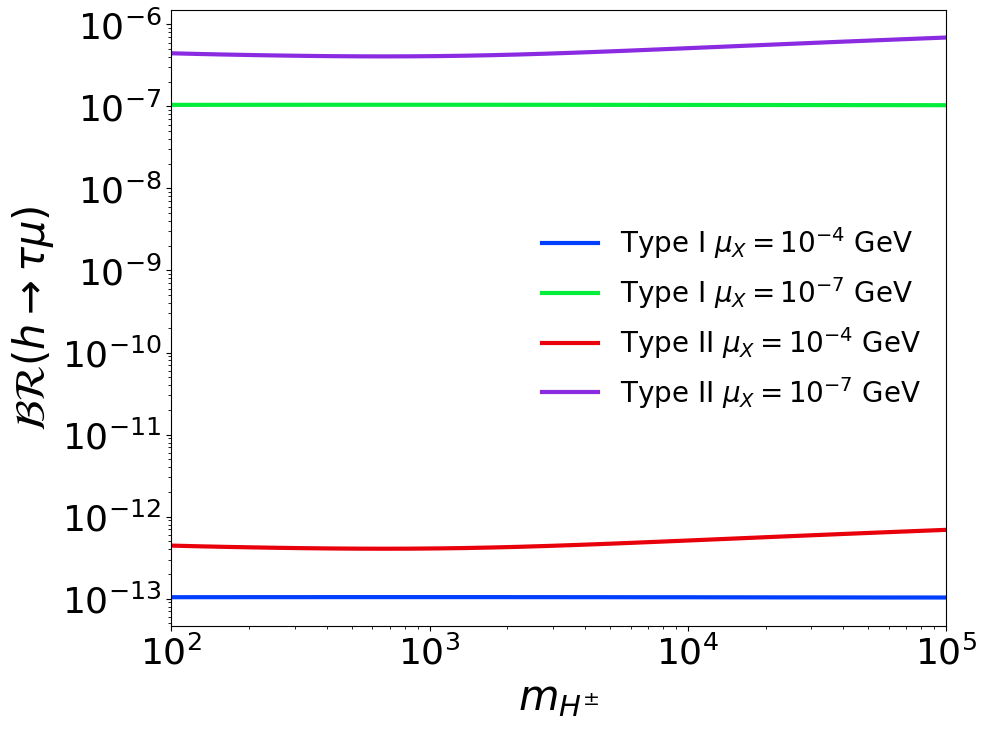}\label{Br:mHpm}}
\subfigure[ ]{\includegraphics[scale = .3]{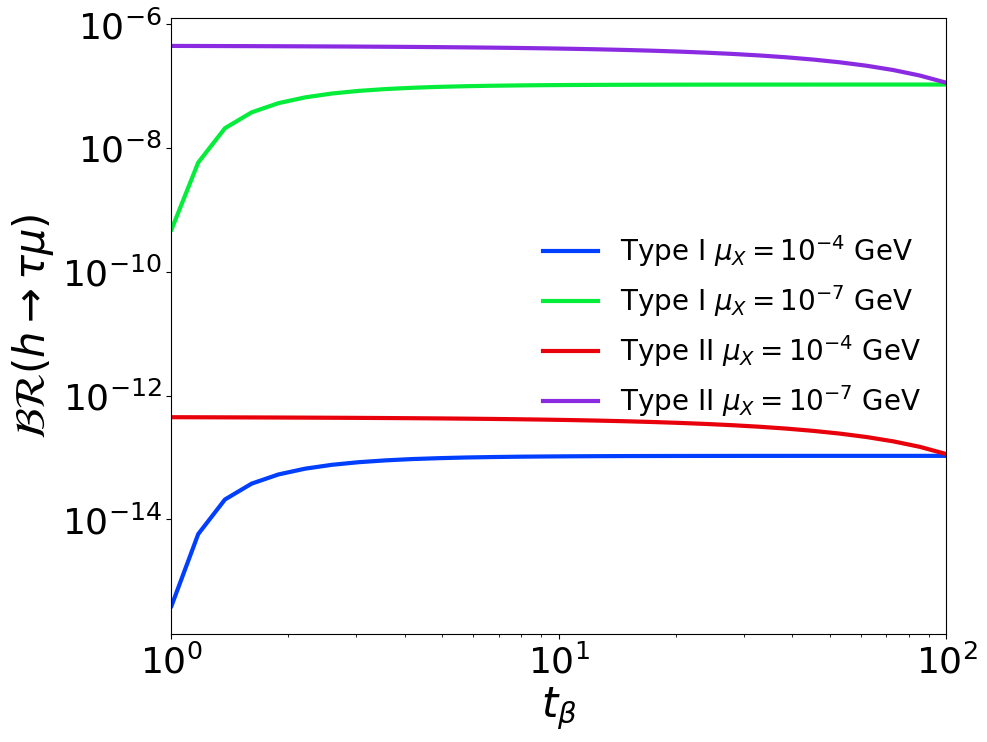}\label{Br:tb}}
\caption{Behavior of branching ratio of $h \to \tau \mu $. In the left panel we show the  $\mathcal{BR}(h \to \tau \mu )$ vs $m_{H^\pm}$ with $t_{\beta} = 10$. In the right panel, we show $\mathcal{BR}(h \to \tau \mu)$ vs $t_{\beta}$ with $m_{H^\pm} = 600$ GeV.}
\label{fig:Brs_2HDM}
\end{figure}

The branching ratio  $\mathcal{BR}(h \to \tau \mu )$ is  shown in Figure~\ref{fig:Brs_2HDM}\footnote{The behavior of $\mathcal{BR}(h \to e  \tau)$ and $\mathcal{BR}(h \to e \mu)$ are similar.}, as function of $t_{\beta}$ and $m_{H^\pm}$ for type I and II of $\nu$2HDM, considering two cases for $\mu_X= 10^{-4}, 10^{-7}$. In Figure~\ref{fig:perturvativityboundtb2}, for the case of $\mu_X= 10^{-4}$ (red dashed line) the perturbativity bound does not constraint the values of $t_{\beta}$, however for the case of $\mu_X= 10^{-7}$ (blue dashed line), the allowed values are $t_{\beta} \gtrsim 5\times 10^{-1}$.
On one hand, in Figure~\ref{Br:mHpm}, we observe that $\mathcal{BR}(h \to \tau \mu )$ is sensible to the scale of $m_{H^\pm}$ for the type II. In contrast to type I, where the dependence on $m_{H^\pm}$ disappear. Only the couplings $h H^\pm G^\mp$ and $h H^\pm H^\mp$ contain explicit dependence on $m_{H^{\pm}}$, see Table~\ref{tab:couplings_2HDM_seesaw}, which appears in diagrams 14, 15 and 16 on Table~\ref{tab:2HDM_contributions}. However, these contributions are not dominant, and as a consequence the weak dependence on $m_{H^{\pm}}$, the difference in type I and II comes from the values of $\xi_h^\ell$ and $\xi_A^\ell$. On the other hand, in Figure~\ref{Br:tb} we observe a dependency over $t_{\beta}$ in both version of $\nu$2HDM. In the type I,  $\mathcal{BR}(h \to \tau \mu )$ is small for $t_{\beta}$ close to 1 and increase with $t_{\beta}$ up to $t_{\beta} \approx 5$, to reach a constant value. Conversely, for type II the behavior is reversed, when $t_{\beta}$ is close to 1, we observe a constant value $\mathcal{BR}(h \to \tau \mu )$ and for values near to $t_{\beta}= 10^2$, the $\mathcal{BR}(h \to \tau \mu )$ decrease. Although, in both models, the neutrinos couple to the same Higgs doublet ($\Phi_2$), the charged leptons do not. The form of $\mathcal{BR}(h \to  \tau \mu )$  in type I and II models shown in Figure~\ref{fig:Brs_2HDM} its on the values of $\xi_h^\ell$ and $\xi_A^\ell$. Lastly, in both models the $\mathcal{BR}(h \to \tau \mu )$ shows a dependency on $\mu_X$ scale, as a consequence of the Casas-Ibarra parametrization~\eqref{ec:ISS_MD}.

\begin{figure}[!htb]
	\centering
	\subfigure[ ]{\includegraphics[scale = 0.22]{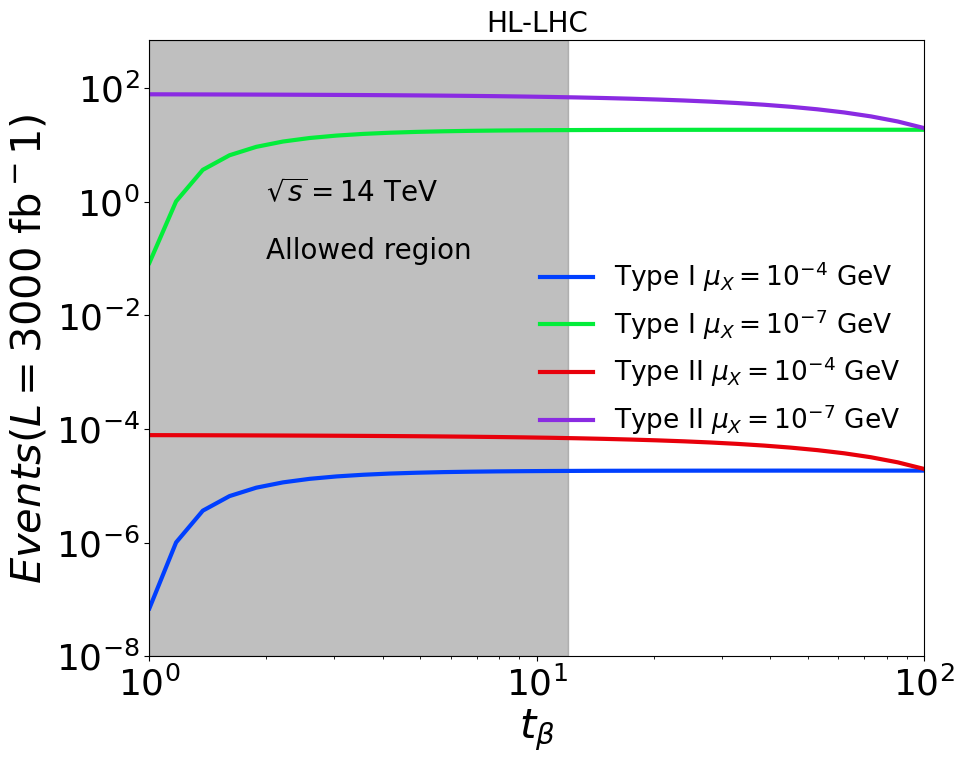}}
	\subfigure[ ]{\includegraphics[scale = 0.22]{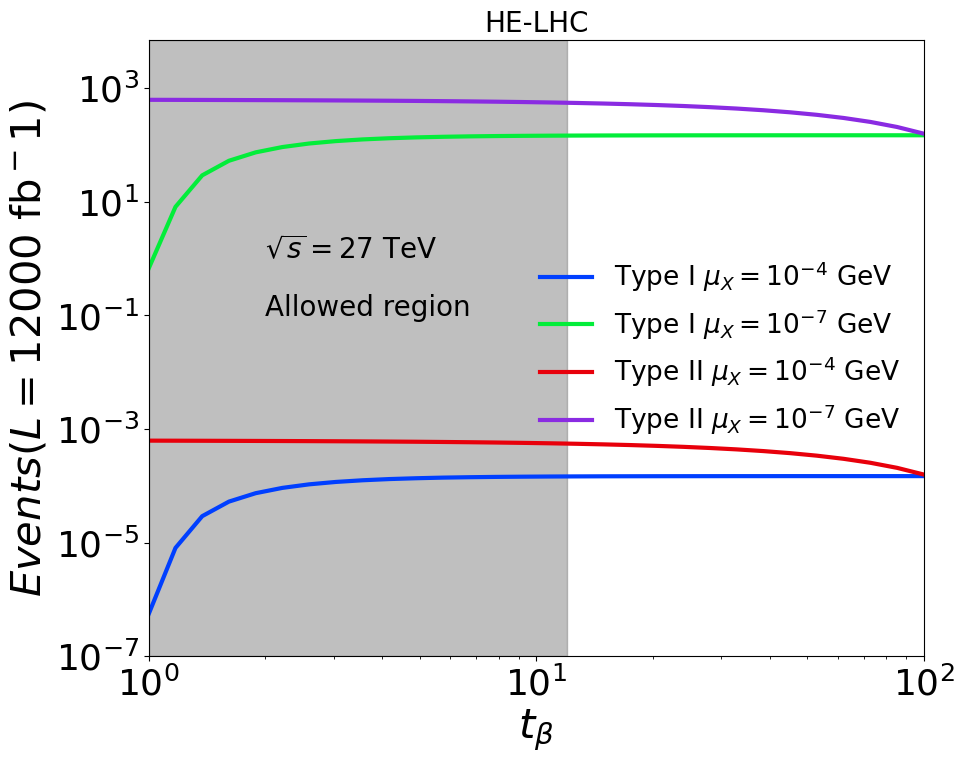}}
	\subfigure[ ]{\includegraphics[scale = 0.22]{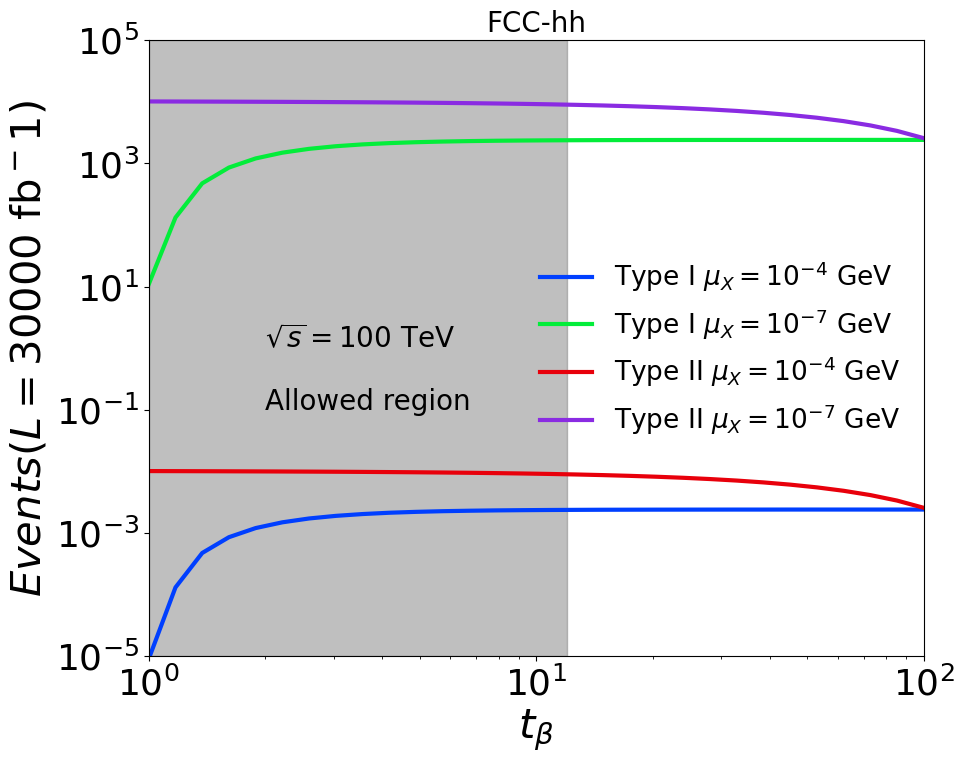}}
	\caption{Events-$t_{\beta}$ plane, for (a) Scenario \textbf{FCH1}, (b) Scenario \textbf{FCH2}, (c) Scenario \textbf{FCH3}.   The dark area corresponds to the allowed region. See Table~\ref{ParValues}.}
	\label{Events_nu2hdm}
\end{figure}

Similarly to Section~\ref{subsec:collideranalysis}, we show the number of  Events-$t_{\beta}$ in the Figure~\ref{Events_nu2hdm}, considering the production cross section  $\sigma(p p \to h \to \tau\mu) = \sigma(gg\to h)\mathcal{BR}(h\to\tau\mu)$ for scenarios $\textbf{FHC1}$, $\textbf{FHC2}$ and $\textbf{FHC3}$. The dark area indicates the allowed region by all observables analyzed previously (see Table~\ref{ParValues}). Given the center-of-mass energy and the integrated luminosity, we find that the maximum number of signal events produced are of the order $\mathcal{N}_{\mathcal{S}}^{\textbf{FHC1}}=\mathcal{O}(10^2)$, $\mathcal{N}_{\mathcal{S}}^{\textbf{FHC2}}=\mathcal{O}(10^3)$, $\mathcal{N}_{\mathcal{S}}^{\textbf{FHC3}}=\mathcal{O}(10^4)$, for $\nu$2HDM-II and $\mu_X = 10^{-7}$ GeV.
%

\section{Conclusions}\label{section:conclusions}
\begin{figure}[!htb]
	\centering
	\includegraphics[scale = 0.35]{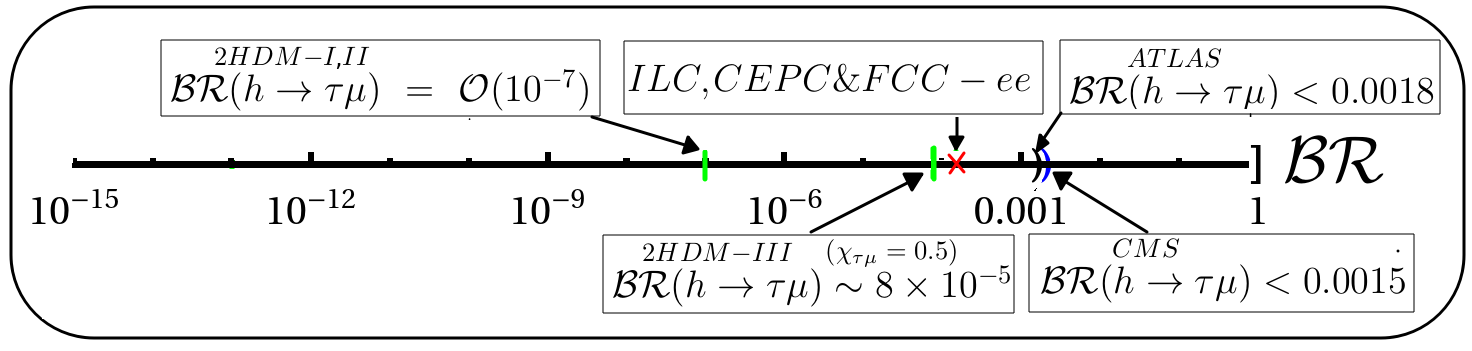}
	\caption{Comparison between experimental results (and projections) and our predictions for the $\mathcal{BR}(h \to \tau\mu)$. We have considered $t_\beta=8$ and $c_{\beta\alpha}=0.01$ in the numerical evaluation. }
	\label{flecha}
\end{figure}

We have presented a general discussion of the search for  the LFV Higgs decays at LHC,  including several versions of the  2HDM, namely: the texturized version (2HDM-Tx), which is a particular version of the general model 2HDM-III, as well as the 2HDM with neutrino masses based on the inverse See-Saw mechanism ($\nu$2HDM).  For the 2HDM-Tx  we started by  deriving the constraints on the model parameters, and the lessons that LHC can give on the model structure. This was followed by the presentation of our study for the limits on $\mathcal{BR}(h \to \tau\mu)$ that can be achieved at LHC, as a function of the integrated luminosities that are projected for the future phases of the LHC.  Then, we also discuss the expected rates for the LFV decay $h \to \ell_a \ell_b$, within the $\nu$2HDM, where such coupling is induced at one-loop level, with some technical details left for the Appendix A. These results are summarized in the Figure~\ref{flecha}. There we can see that current limits from ATLAS and CMS on  $\mathcal{BR}(h \to \tau\mu)$ are getting quite close to the value $10^{-3}$. Thus, values of the parameter $\chi_{\tau\mu}$ of order $1$ $(0.5)$ are excluded for values of $t_\beta$ larger than about $20$ $(35)$. However, we can also see that for $t_\beta=8$, $c_{\alpha\beta}=0.01$, values of $\chi_{\tau\mu}=0.5$ imply values of $\mathcal{BR}(h \to \tau\mu)$ about $10^{-4}$, which are still consistent with current limits, but could be rule out in searches for future eletron-positron colliders. 

On the other hand, for the loop-induced LFV Higgs decay, we find that  $\mathcal{BR}(h \to \tau\mu)$ is about $10^{-7}$ for the version $\nu$2HDM-I, and $10^{-13}$ for the version  $\nu$2HDM-II, which are still far from the experimental reach of LHC and future colliders.

Thus, we have learned from LHC that models with large tree-level LFV Higgs couplings, as arising in the 2HDM-III, are being constrained by current data, but even the model version with textures (2HDM-Tx) has regions of parameters that pass current constraints. The coming phases of LHC with higher luminosity will be able to probe models where such LFV Higgs couplings arise at tree-level, but include a suppression mechanism, as in MFV or BGL models~\cite{Branco:1996bq}. Finally, we conclude that models where such couplings arise at loop-level, as in the MSSM or 2HDM with neutrino masses, seem to be out of the reach of LHC14.

{\bf{Acknowledgments}} 

The work of Marco A. Arroyo-Ure\~na is supported by ``Estancias posdoctorales por M\'exico (CONAHCYT)" and ``Sistema Nacional de Investigadores" (SNI-CONAHCYT). JLD-C and O.F.B. acknowledge the support of SNI-CONAHCYT and VIEP (BUAP). M. Zeleny thanks to ESTANCIA POSDOCTORALES REALIZADA GRACIAS AL PROGRAMA DE BECAS POSDOCTORALES EN LA UNAM(POSDOC) for postdoctoral funding.
\appendix

\section{Scalar couplings for LFV-HD and form factors}\label{appen:FormFactors2HDM}
We present the couplings of SM-like Higgs and charged currents in Table~\ref{tab:couplings_2HDM_seesaw}. In this case we use the chiral notation in agreement with~\cite{Zeleny-Mora:2021tym},
\begin{table}[!htb]
	\centering
	\scalebox{0.90}{
		\begin{tabular}{cccc}
			\hline \hline
			\rule[-1ex]{0pt}{2.5ex} {\bf Vertex}& {\bf Coupling} & {\bf Vertex} & {\bf Coupling}\\
			\hline \hline
			\rule[-1ex]{0pt}{2.5ex} $h W^+ W^-$ & $ig m_W \Xi_h$ &
			$h G^+ G^-$ & $-i \dfrac{g m_h^2 \Xi_h}{2 m_W}$ \\
			\hline
			\rule[-1ex]{0pt}{2.5ex} $h G^+ W^-$ & $i \dfrac{g \Xi_h}{2}(p_+ - p_0)_\mu$ & $h W^+ G^-$ & $i \dfrac{g \Xi_h}{2}(p_0 - p_-)_\mu$ \\
			\hline
			\rule[-1ex]{0pt}{2.5ex} $h H^+ W^-$ & $i \dfrac{g \eta_h}{2}(k_+ - p_0)_\mu$ &
			$h W^+ H^-$ & $i \dfrac{g \eta_h}{2 }(p_0 - k_-)_\mu$\\
			\hline
			\rule[-1ex]{0pt}{2.5ex} $h H^{\pm}G^{\mp}$ & $i \dfrac{g \eta_h(m_{H^{\pm}}^2 - m_h^2)}{2 m_W}$ & $h H^{\pm}H^{\mp}$ & $ig \dfrac{\rho_h \mathcal{K}_h - \Delta_h \mathcal{Q}_h}{4 m_W \sin{2 \beta}} + i \dfrac{4 \lambda_5  m_W \rho_h}{g \sin{2 \beta}}$ \\
			\hline
			\rule[-1ex]{0pt}{2.5ex} $h l \overline{l}$ & $-ig c_h^{l}\dfrac{m_l}{2 m_W}$ & $h n_i n_j$ &
			$\dfrac{-i g c^n_h}{2 m_W}\left[C_{i j}\left(P_{L} m_{n_{i}}+P_{R} m_{n_{j}}\right) + C_{i j}^{*}\left(P_{L} m_{n_{j}}+P_{R} m_{n_{i}}\right)\right]$ \\
			\hline
			\rule[-1ex]{0pt}{2.5ex} $\bar{n}_{i} \ell_{a} W_{\mu}^{+}$ & $\dfrac{i g}{\sqrt{2}} U_{a i}^{\nu} \gamma^{\mu}P_{L}$ & $\overline{\ell_{a}} n_{j} W_{\mu}^{-}$ & $\dfrac{i g}{\sqrt{2}} U_{a j}^{\nu *} \gamma^{\mu} P_{L}$ \\
			\hline
			\rule[-1ex]{0pt}{2.5ex} $\bar{n}_{i} \ell_{a} G_{W}^{+}$ & $-\dfrac{i g}{\sqrt{2} m_{W}} U_{a i}^{\nu}\left(m_{\ell_{a}} P_{R}-m_{n_i} P_{L}\right)$ & $\overline{\ell_{a}} n_{j} G_{W}^{-}$ & $-\dfrac{i g}{\sqrt{2} m_{W}} U_{a j}^{\nu*}\left(m_{\ell_{a}} P_{L}-m_{n_j} P_{R}\right)$ \\
			\hline
			\rule[-1ex]{0pt}{2.5ex} $\bar{n}_{i} \ell_{a} H^{+}$ & $-\dfrac{i g }{\sqrt{2} m_W}U^{\nu}_{a i}(e_{A}^{l}m_{\ell_a} P_R -e_{A}^{n}m_{n_i} P_L)$ & $\overline{\ell_{a}} n_{j} H^{-}$ & $-\dfrac{i g}{\sqrt{2} m_W} U^{\nu *}_{a j}(  e_{A}^{l}m_{\ell_a}P_L - e_{A}^{n}m_{n_i} P_R)$ \\
			\hline \hline
	\end{tabular}	}
	\caption{Couplings of $h$, $G^{\pm}$, $W^{\pm}$ and $H^{\pm}$, for LFV-HD in the $\nu$2HDM-I, II.}
	\label{tab:couplings_2HDM_seesaw}
\end{table}
with the following definitions:
\begin{center}
	\begin{align}
	\begin{split}
	\Xi_h  &= \sin{\beta - \alpha}, \\
	\mathcal{K}_h &= 4 m_A^2 -3 m_h^2-2 m_{H^\pm}^2, \\
	\rho_h &= \cos{\alpha + \beta},
	\end{split}
	&& 
	\begin{split}
	\eta_h &= \cos{\beta - \alpha}, \\
	\mathcal{Q}_h &= m_h^2 - 2 m_{H^\pm}^2,  \\
	\Delta_h &= \cos{\alpha -  3\beta}.
	\end{split}
	\end{align}
\end{center}
In the special case of the $\nu$2HDM-I, II, 18 Feynman graphs contributes to LFV-HD at one loop,  summarized in Table~\ref{tab:2HDM_contributions}, following the topologies shown in Figure~\ref{fig:FeynmanDiagrams}.
\begin{center}
	{
		\begin{table}
			\centering
			\begin{tabular}{cccccccccccc}
				\hline \hline
				\rule[-1ex]{0pt}{2.5ex}{\bf No.} & {\bf Structure}  &{\bf Diagram} & $P_0$ & $P_1$ & $P_2$ & \rule[-1ex]{0pt}{2.5ex}{\bf No.} & {\bf Structure}  &{\bf Diagram} & $P_0$ & $P_1$ & $P_2$  \\
				\hline \hline
				\rule[-1ex]{0pt}{2.5ex} 1& SFF & \ref{fig:FeynmanDiagrams}(i)& $G_W$ & $\overline{n}_i$ & $n_j$ & \rule[-1ex]{0pt}{2.5ex} 11& SFF & \ref{fig:FeynmanDiagrams}(i) & $H^{\pm}$  & $\overline{n}_i$ &  $n_j$ \\
				\hline
				\rule[-1ex]{0pt}{2.5ex} 2& VFF & \ref{fig:FeynmanDiagrams}(j)& $W$ &$\overline{n}_i$ &  $n_j$ & \rule[-1ex]{0pt}{2.5ex} 12& FVS & \ref{fig:FeynmanDiagrams}(c) & $n_i$  &$W$ &  $H^{\pm}$ \\
				\hline
				\rule[-1ex]{0pt}{2.5ex} 3& FSS & \ref{fig:FeynmanDiagrams}(a)& $n_i$  &$G_W$ &  $G_W$ & \rule[-1ex]{0pt}{2.5ex} 13& FSV & \ref{fig:FeynmanDiagrams}(b) & $n_i$  &$H^{\pm}$ &  $W$  \\
				\hline
				\rule[-1ex]{0pt}{2.5ex} 4& FVS & \ref{fig:FeynmanDiagrams}(c) & $n_i$  &$W$ &  $G_W$ & \rule[-1ex]{0pt}{2.5ex} 14& FSS & \ref{fig:FeynmanDiagrams}(a)& $n_i$  &$G_W$ &  $H^{\pm}$  \\
				\hline
				\rule[-1ex]{0pt}{2.5ex} 5& FSV & \ref{fig:FeynmanDiagrams}(b) & $n_i$  &$G_W$ &  $W$  & \rule[-1ex]{0pt}{2.5ex} 15& FSS & \ref{fig:FeynmanDiagrams}(a) & $n_i$  &$H^{\pm}$ &  $G_W$  \\
				\hline
				\rule[-1ex]{0pt}{2.5ex} 6& FVV & \ref{fig:FeynmanDiagrams}(d) & $n_i$  &$W$ &  $W$ & \rule[-1ex]{0pt}{2.5ex} 16& FSS & \ref{fig:FeynmanDiagrams}(a) & $n_i$  &$H^{\pm}$ &  $H^{\pm}$  \\
				\hline
				\rule[-1ex]{0pt}{2.5ex} 7& FV & \ref{fig:FeynmanDiagrams}(g) & $n_i$ &$W$ & --- & \rule[-1ex]{0pt}{2.5ex} 17& FS & \ref{fig:FeynmanDiagrams}(e)& $n_i$  &$H^{\pm}$ &  ---  \\
				\hline
				\rule[-1ex]{0pt}{2.5ex} 8 &FS & \ref{fig:FeynmanDiagrams}(e) & $n_i$  &$G_W$ &  ---  & \rule[-1ex]{0pt}{2.5ex} 18& SF & \ref{fig:FeynmanDiagrams}(f) & $n_i$  & --- &  $H^{\pm}$ \\
				\hline
				\rule[-1ex]{0pt}{2.5ex} 9& VF & \ref{fig:FeynmanDiagrams}(h) & $n_i$  & ---  & $W$ & ---& --- & --- & --- & --- &  --- \\
				\hline
				\rule[-1ex]{0pt}{2.5ex} 10& SF & \ref{fig:FeynmanDiagrams}(f) & $n_i$  & --- &  $G_W$ & ---& --- & ---& ---  & --- &  --- \\ \hline \hline
			\end{tabular}
			\caption{Particles involved in each one loop diagram that contribute to  $\phi \to \ell_a^+ \ell_b^-$ in the $\nu$2HDM.}
			\label{tab:2HDM_contributions}
		\end{table}
	}
\end{center}
 The associated form factors are as follows~\cite{Zeleny-Mora:2021tym},
\begin{equation}\label{ec:Dijab_2HDM}
\Delta_{ij}^{ab} = \dfrac{g^{3}}{64 \pi^{2} m_{W}^{3}} U_{bj}^{\nu} U_{ai}^{\nu *}.
\end{equation}
\begin{align}\notag
A_L^{(1)}(G\overline{n}_i n_j) &= {m}_{\ell_a} c^n_\phi \sum_{i,j=1}^{K+3}  \left[\left(\left(\operatorname{{{B^{(12)}_{0}}}} + m_{W}^{2} \operatorname{C_{0}}\right) m_{n_j}^{2} - \left({m}_{\ell_a}^{2} m_{n_j}^{2} + {m}_{\ell_b}^{2} m_{n_i}^{2} - 2 m_{n_i}^{2} m_{n_j}^{2}\right) \operatorname{C_{1}} \right) {C}_{ij}  \right.\\ \notag & \left. 
+ \left(\operatorname{{{B^{(12)}_{0}}}} + m_{W}^{2} \operatorname{C_{0}} - \left({m}_{\ell_a}^{2} + {m}_{\ell_b}^{2} - m_{n_i}^{2} - m_{n_j}^{2}\right) \operatorname{C_{1}} \right) {{{C^*}}}_{ij} m_{n_i} m_{n_j}\right] \Delta_{ij}^{ab} , \\ \notag
A_R^{(1)}(G\overline{n}_i n_j) & =  {m}_{\ell_b} c^n_\phi \sum_{i,j = 1}^{K+3} \left[\left(\left(\operatorname{{{B^{(12)}_{0}}}} + m_{W}^{2} \operatorname{C_{0}}\right) m_{n_i}^{2} + \left({m}_{\ell_a}^{2} m_{n_j}^{2} + {m}_{\ell_b}^{2} m_{n_i}^{2} - 2 m_{n_i}^{2} m_{n_j}^{2}\right) \operatorname{C_{2}} \right) {C}_{ij} \right.\\ & \left. + \left(\operatorname{{{B^{(12)}_{0}}}} + m_{W}^{2} \operatorname{C_{0}} + \left({m}_{\ell_a}^{2} + {m}_{\ell_b}^{2} - m_{n_i}^{2} - m_{n_j}^{2}\right) \operatorname{C_{2}} \right) {{{C^*}}}_{ij} m_{n_i} m_{n_j} \right]  \Delta_{ij}^{ab}.
\end{align}
\begin{align}\notag
A_L^{(2)}(W\overline{n}_i n_j) &= 2 m_{W}^{2} {m}_{\ell_a} c^n_\phi \sum_{i,j = 1}^{K+3}  \left(\left(\left(m_{n_i}^{2} + m_{n_j}^{2}\right) \operatorname{C_{1}} - \operatorname{C_{0}} m_{n_j}^{2}\right) {C}_{ij} - \left(\operatorname{C_{0}} - 2 \operatorname{C_{1}}\right) {{{C^*}}}_{ij} m_{n_i} m_{n_j}\right)  \Delta_{ij}^{ab}, \\
A_R^{(2)}(W\overline{n}_i n_j) & =- 2 m_{W}^{2} {m}_{\ell_b} c^n_\phi \sum_{i,j = 1}^{K+3} \left(\left(\left(m_{n_i}^{2} + m_{n_j}^{2}\right) \operatorname{C_{2}} + \operatorname{C_{0}} m_{n_i}^{2}\right) {C}_{ij} + \left(\operatorname{C_{0}} + 2 \operatorname{C_{2}}\right) {{{C^*}}}_{ij} m_{n_i} m_{n_j}\right) \Delta_{ij}^{ab} . 
\end{align}
\begin{align}\notag
A_L^{(3)}(n_i W^+ W^-) & = - 4 m_{W}^{4} {m}_{\ell_a}  \Xi_\phi \sum_{i = 1}^{K+3}\operatorname{C_{1}}\Delta_{ii}^{ab},\\
A_R^{(3)}(n_i W^+ W^-) & =  4 m_{W}^{4}  {m}_{\ell_b}  \Xi_\phi \sum_{i = 1}^{K+3}\operatorname{C_{2}}\Delta_{ii}^{ab}.
\end{align}
\begin{align}\notag 
A_L^{(4)}(n_i W^- G^+) 
& = m_{W}^{2}{m}_{\ell_a} \Xi_\phi \sum_{i=1}^{K+3} \left[\left(2 {m}_{\ell_b}^{2} - m_{n_i}^{2}\right) \operatorname{C_{1}} - \operatorname{C_{0}} m_{n_i}^{2} - \operatorname{C_{2}} {m}_{\ell_b}^{2}\right] \Delta_{ii}^{ba}, \\ 
A_R^{(4)}(n_i W^- G^+) &=  m_{W}^{2} {m}_{\ell_b} \Xi_\phi \sum_{i=1}^{K+3} \left[\operatorname{{{B^{(12)}_{0}}}} + \left(2 {m}_{\ell_b}^{2} + m_{n_i}^{2}\right) \operatorname{C_{2}} - \left({m}_{\ell_a}^{2} + 2 {m}_{\ell_b}^{2} - 2 {m}_{h}^{2}\right) \operatorname{C_{1}} + 3 \operatorname{C_{0}} m_{n_i}^{2} \right] \Delta_{ii}^{ba},
\end{align}
\begin{align}\notag
A_L^{(5)}(n_i G^- W^+) & =  m_{W}^{2} m_{\ell_a} \Xi_\phi \sum_{i=1}^{K+3}m_{n_i}  \left(\operatorname{{{B^{(12)}_{0}}}}- \left(2 {m}_{\ell_a}^{2} + m_{n_i}^{2}\right) \operatorname{C_{1}} + \left(2 {m}_{\ell_a}^{2} + {m}_{\ell_b}^{2} - 2 {m}_{h}^{2}\right) \operatorname{C_{2}} + 3 \operatorname{C_{0}} m_{n_i}^{2} \right)  \Delta_{ii}^{ab},   \\
A_R^{(5)}(n_i G^- W^+) & =  m_{W}^{2} {m}_{\ell_b} \Xi_\phi \sum_{i=1}^{K+3}m_{n_i}  \left(- \left(2 {m}_{\ell_a}^{2} - m_{n_i}^{2}\right) \operatorname{C_{2}} - \operatorname{C_{0}} m_{n_i}^{2} + \operatorname{C_{1}} {m}_{\ell_a}^{2}\right) \Delta_{ii}^{ab} .
\end{align}
\begin{align}\notag
A_L^{(6)}(n_i G^- G^+) & = m_{\ell_a} {m}_{\phi}^{2} \Xi_\phi \sum_{i=1}^{K+3} \left(\left( \operatorname{C_{0}}  - \operatorname{C_{1}} \right) m_{n_i}^{2} + \operatorname{C_{2}} {m}_{\ell_b}^{2}\right) \Delta_{ii}^{ab},  \\
A_R^{(6)}(n_i G^- G^+) & = m_{\ell_b} {m}_{\phi}^{2} \Xi_\phi \sum_{i=1}^{K+3} \left(\left( \operatorname{C_{0}} +  \operatorname{C_{2}} \right) m_{n_i}^{2} - \operatorname{C_{1}} {m}_{\ell_a}^{2}\right) \Delta_{ii}^{ab} . 
\end{align}
\begin{align}\notag
A_L^{(7)}(n_i W) & = - \dfrac{2 m_{W}^{2}  {m}_{\ell_a} {m}_{\ell_b}^{2} }{{m}_{\ell_a}^{2} - {m}_{\ell_b}^{2}} c^l_\phi  \sum_{i=1}^{K+3}\operatorname{{{B^{(1)}_{1}}}}\Delta_{ii}^{ab} , \\
A_R^{(7)}(n_i W) & = - \dfrac{2 m_{W}^{2}  {m}_{\ell_a}^2 {m}_{\ell_b} }{{m}_{\ell_a}^{2} - {m}_{\ell_b}^{2}}
c^l_\phi  \sum_{i=1}^{K+3}\operatorname{{{B^{(1)}_{1}}}}\Delta_{ii}^{ab}. 
\end{align}
\begin{align}\notag
A_L^{(8)}(n_i G) & =   \dfrac{{m}_{\ell_a} {m}_{\ell_b}^2 }{{m}_{\ell_a}^{2} - {m}_{\ell_b}^{2}} c^l_\phi  \sum_{i=1}^{K+3}  \left(- \left({m}_{\ell_a}^{2} + m_{n_i}^{2}\right) \operatorname{{{B^{(1)}_{1}}}} + 2 \operatorname{{{B^{(1)}_{0}}}} m_{n_i}^{2}\right) \Delta_{ii}^{ab},\\
A_R^{(8)}(n_i G) & =  \dfrac{ {m}_{\ell_b} }{{m}_{\ell_a}^{2} - {m}_{\ell_b}^{2}} c^l_\phi  \sum_{i=1}^{K+3} \left(\left({m}_{\ell_a}^{2} + {m}_{\ell_b}^{2}\right) \operatorname{{{B^{(1)}_{0}}}} m_{n_i}^{2} - \left({m}_{\ell_b}^{2} + m_{n_i}^{2}\right) \operatorname{{{B^{(1)}_{1}}}} {m}_{\ell_a}^{2}\right)\Delta_{ii}^{ab}.
\end{align}
\begin{align}\notag
A_L^{(9)}(W n_i) & = - \dfrac{2 m_{W}^{2}  {m}_{\ell_a} {m}_{\ell_b}^{2} }{ {m}_{\ell_a}^{2} - {m}_{\ell_b}^{2}} c^l_\phi  \sum_{i=1}^{K+3}\operatorname{{{B^{(2)}_{1}}}}\Delta_{ii}^{ab},\\
A_R^{(9)}(W n_i) & = - \dfrac{2 m_{W}^{2}  {m}_{\ell_a}^{2} {m}_{\ell_b} }{ {m}_{\ell_a}^{2} - {m}_{\ell_b}^{2}}
c^l_\phi  \sum_{i=1}^{K+3}\operatorname{{{B^{(2)}_{1}}}}\Delta_{ii}^{ab}.
\end{align}
\begin{align}\notag
A_L^{(10)}(G n_i) & = - \dfrac{ {m}_{\ell_a}}{{m}_{\ell_a}^{2} - {m}_{\ell_b}^{2}} c^l_\phi  \sum_{i=1}^{K+3} \left(\left({m}_{\ell_a}^{2} + {m}_{\ell_b}^{2}\right) \operatorname{{{B^{(2)}_{0}}}} m_{n_i}^{2} + \left({m}_{\ell_a}^{2} + m_{n_i}^{2}\right) \operatorname{{{B^{(2)}_{1}}}} {m}_{\ell_b}^{2}\right) \Delta_{ii}^{ab}, \\
A_R^{(10)}(G n_i) & = - \dfrac{ {m}_{\ell_a}^2 {m}_{\ell_b}}{{m}_{\ell_a}^{2} - {m}_{\ell_b}^{2}} c^l_\phi  \sum_{i=1}^{K+3}\left(\left({m}_{\ell_b}^{2} + m_{n_i}^{2}\right) \operatorname{{{B^{(2)}_{1}}}} + 2 \operatorname{{{B^{(2)}_{0}}}} m_{n_i}^{2}\right)\Delta_{ii}^{ab}.
\end{align}
\begin{align}\notag
A_L^{(11)}(H^{\pm}\overline{n}_i n_j) &=  - {m}_{a} c^n_\phi  \sum_{i,j=1}^{K+3}  \left[\left( K_1^L B^{(12)}_0 + K_2^L \operatorname{C}_0 + K_3^L \operatorname{C}_1 + K_4^L \operatorname{C}_2\right) {C}_{ij}  
\right.\\ & \left. 
+ \left( Q_1^L B^{(12)}_0 + Q_2^L \operatorname{C}_0 + Q_3^L \operatorname{C}_1 + Q_4^L \operatorname{C}_2 \right) {{{C^*}}}_{ij} m_{n_i} m_{n_j}\right] \Delta_{ij}^{ab} , \\ \notag
A_R^{(11)}(H^{\pm}\overline{n}_i n_j) & =  {m}_{b} c^n_\phi  \sum_{i,j = 1}^{K+3} \left[\left( K_1^R B^{(12)}_0 + K_2^R \operatorname{C}_0 + K_3^R \operatorname{C}_1 + K_4^R \operatorname{C}_2 \right) {C}_{ij} \right.\\ & \left. + \left(Q_1^R B^{(12)}_0 + Q_2^R \operatorname{C}_0 + Q_3^R \operatorname{C}_1 + Q_4^R \operatorname{C}_2 \right) {{{C^*}}}_{ij} m_{n_i} m_{n_j} \right]  \Delta_{ij}^{ab},
\end{align}
where $K^{L}_{k}$ and $Q^{L}_{k}$ with  $k= 1,\dots ,4$, are given by
%
\begin{align}
K_1^L &= - {{e^{l}_{A}}} {{e^{n}_{A}}} {{{m_n}}}_{j}^{2},\\
K_2^L &= \displaystyle \left({{e^{l}_{A}}}\right)^{2} {m}_{\ell_b}^{2} {{{m_n}}}_{i}^{2} - {{e^{l}_{A}}} {{e^{n}_{A}}} \left(m_{H^{\pm}}^2 {{{m_n}}}_{j}^{2} + {m}_{\ell_b}^{2} {{{m_n}}}_{i}^{2} + {{{m_n}}}_{i}^{2} {{{m_n}}}_{j}^{2}\right) + \left({{e^{n}_{A}}}\right)^{2} {{{m_n}}}_{i}^{2} {{{m_n}}}_{j}^{2},\\ 
K_3^L &= {{e^{n}_{A}}} \left({{e^{l}_{A}}} \left({m}_{\ell_a}^{2} {{{m_n}}}_{j}^{2} + {m}_{\ell_b}^{2} {{{m_n}}}_{i}^{2}\right) - 2 {{e^{n}_{A}}} {{{m_n}}}_{i}^{2} {{{m_n}}}_{j}^{2}\right),\\
K_4^L &= {{e^{l}_{A}}} \left({{e^{l}_{A}}} - {{e^{n}_{A}}}\right) \left({{{m_n}}}_{i}^{2} + {{{m_n}}}_{j}^{2}\right) {m}_{\ell_b}^{2},\\
Q^1_L &= - {{e^{l}_{A}}} {{e^{n}_{A}}},\\
Q^2_L &= \left({{e^{l}_{A}}}\right)^{2} {m}_{\ell_b}^{2} - {{e^{l}_{A}}} {{e^{n}_{A}}} \left(m_{H^{\pm}}^2 + {m}_{\ell_b}^{2} + {{{m_n}}}_{j}^{2}\right) + \left({{e^{n}_{A}}}\right)^{2} {{{m_n}}}_{j}^{2},\\ 
Q^3_L &= {{e^{n}_{A}}} \left({{e^{l}_{A}}} \left({m}_{\ell_a}^{2} + {m}_{\ell_b}^{2}\right) - {{e^{n}_{A}}} \left({{{m_n}}}_{i}^{2} + {{{m_n}}}_{j}^{2}\right)\right), \\
Q^4_L &= 2 {{e^{l}_{A}}} \left({{e^{l}_{A}}} - {{e^{n}_{A}}}\right) {m}_{\ell_b}^{2},
\end{align}
also, $K^{R}_{k}$ and $Q^{R}_{k}$ are
%
%
\begin{align}
K_1^R &= - K_1^L \dfrac{{{{m_n}}}_{i}^{2}}{{{{m_n}}}_{j}^{2}},\\
K_2^R &= - \left({{e^{l}_{A}}}\right)^{2} {m}_{\ell_a}^{2} {{{m_n}}}_{j}^{2} + {{e^{l}_{A}}} {{e^{n}_{A}}} \left(m_{H^{\pm}}^2 {{{m_n}}}_{i}^{2} + {m}_{\ell_a}^{2} {{{m_n}}}_{j}^{2} + {{{m_n}}}_{i}^{2} {{{m_n}}}_{j}^{2}\right) - \left({{e^{n}_{A}}}\right)^{2} {{{m_n}}}_{i}^{2} {{{m_n}}}_{j}^{2}, \\ 
K_3^R &= K_4^L \dfrac{m_{\ell_a}^2}{m_{\ell_b}^2},\\
K_4^R &= K_3^L,\\
Q_1^R &= - Q^1_L, \\
Q_2^R &= - \left({{e^{l}_{A}}}\right)^{2} {m}_{\ell_a}^{2} + {{e^{l}_{A}}} {{e^{n}_{A}}} \left(m_{H^{\pm}}^2 + {m}_{\ell_a}^{2} + {{{m_n}}}_{i}^{2}\right) - \left({{e^{n}_{A}}}\right)^{2} {{{m_n}}}_{i}^{2}, \\  Q_3^R &=Q^4_L \dfrac{m_{\ell_a}^2}{m_{\ell_b}^2}.\\
Q_4^R &= Q^3_L
\end{align}
\begin{align} 
A_L^{(12)}(n_i W^- H^+) 
& = m_{W}^{2}{m}_{\ell_a} \eta_\phi \sum_{i=1}^{K+3} \left[{{e^{l}_{A}}} \operatorname{C_{2}} {m}_{\ell_b}^{2} + {{e^{n}_{A}}} \operatorname{C_{0}} {{{m_n}}}_{i}^{2} - \left(2 {{e^{l}_{A}}} {m}_{\ell_b}^{2} - {{e^{n}_{A}}} {{{m_n}}}_{i}^{2}\right) \operatorname{C_{1}}\right] \Delta_{ii}^{ba}, \\ \notag
A_R^{(12)}(n_i W^- H^+) &=  m_{W}^{2} {m}_{\ell_b} \eta_\phi \sum_{i=1}^{K+3} \left[{{e^{l}_{A}}} \left(- 2 {m}_{\phi}^{2} + {m}_{\ell_a}^{2} + 2 {m}_{\ell_b}^{2}\right) \operatorname{C_{1}} - {{e^{l}_{A}}} \operatorname{{{B^{(12)}_{0}}}} 
\right.\\ & \left. \qquad
- \left({{e^{l}_{A}}} + 2 {{e^{n}_{A}}}\right) \operatorname{C_{0}} {{{m_n}}}_{i}^{2} - \left(2 {{e^{l}_{A}}} {m}_{\ell_b}^{2} + {{e^{n}_{A}}} {{{m_n}}}_{i}^{2}\right) \operatorname{C_{2}} \right] \Delta_{ii}^{ba},
\end{align}
\begin{align}\notag
A_L^{(13)}(n_i H^- W^+) & =  m_{W}^{2} m_{\ell_a} \eta_\phi \sum_{i=1}^{K+3}  \left[\left(2 {{e^{l}_{A}}} {m}_{\ell_a}^{2} + {{e^{n}_{A}}} {{{m_n}}}_{i}^{2}\right) \operatorname{C_{1}} - {{e^{l}_{A}}} \left(- 2 {m}_{\phi}^{2} + 2 {m}_{\ell_a}^{2} + {m}_{\ell_b}^{2}\right) \operatorname{C_{2}} 
\right.\\ & \left. \qquad
- {{e^{l}_{A}}} \operatorname{{{B^{(12)}_{0}}}} - \left({{e^{l}_{A}}} + 2 {{e^{n}_{A}}}\right) \operatorname{C_{0}} {{{m_n}}}_{i}^{2}\right]  \Delta_{ii}^{ab},   \\
A_R^{(13)}(n_i H^- W^+) & =  m_{W}^{2} {m}_{\ell_b} \eta_\phi \sum_{i=1}^{K+3}  \left[{{e^{n}_{A}}} \operatorname{C_{0}} {{{m_n}}}_{i}^{2} - {{e^{l}_{A}}} \operatorname{C_{1}} {m}_{\ell_a}^{2} + \left(2 {{e^{l}_{A}}} {m}_{\ell_a}^{2} - {{e^{n}_{A}}} {{{m_n}}}_{i}^{2}\right) \operatorname{C_{2}}\right] \Delta_{ii}^{ab} .
\end{align}
\begin{align}\notag 
A_L^{(14)}(n_i G^- H^+) 
& = {m}_{\ell_a} \eta_\phi \sum_{i=1}^{K+3} \left[{{e^{l}_{A}}} \left({m}_{\phi}^{2} - \left({{m_{H^{\pm}}}}\right)^{2} \right) \operatorname{C_{2}} {m}_{\ell_b}^{2} 
\right.\\ & \left. \qquad
+ {{e^{n}_{A}}} \left(\left( {m}_{\phi}^{2}- \left({{m_{H^{\pm}}}}\right)^{2} \right)  \left( \operatorname{C_{0}} - \operatorname{C_{1}} \right)   \right) {{{m_n}}}_{i}^{2}\right] \Delta_{ii}^{ba}, \\ \notag
A_R^{(14)}(n_i G^- H^+) &=  {m}_{\ell_b} \eta_\phi \sum_{i=1}^{K+3} \left[ - {{e^{l}_{A}}} \left({m}_{\phi}^{2} - \left({{m_{H^{\pm}}}}\right)^{2} \right) \operatorname{C_{1}} {m}_{\ell_a}^{2} 
\right.\\ & \left. \qquad
+ \left( \left({{e^{l}_{A}}} \operatorname{C_{0}} - {{e^{n}_{A}}} \operatorname{C_{1}} \right) \left(  {m}_{\phi}^{2} - \left({{m_{H^{\pm}}}}\right)^{2} \right) \right) {{{m_n}}}_{i}^{2} \right] \Delta_{ii}^{ba},
\end{align}
\begin{align}\notag
A_L^{(15)}(n_i H^- G^+) & =  m_{\ell_a} \eta_\phi \sum_{i=1}^{K+3}  \left[ {{e^{l}_{A}}} \left({m}_{\phi}^{2} - \left({{m_{H^{\pm}}}}\right)^{2} \right) \operatorname{C_{2}} {m}_{\ell_b}^{2} 
\right.\\ & \left. \qquad
+ \left( \left( {{e^{l}_{A}}} \operatorname{C_{0}} - {{e^{n}_{A}}} \operatorname{C_{1}} \right)  \left({m}_{\phi}^{2}- \left({{m_{H^{\pm}}}}\right)^{2} \right)  \right) {{{m_n}}}_{i}^{2} \right]  \Delta_{ii}^{ab},   \\ \notag
A_R^{(15)}(n_i H^- G^+) & =  {m}_{\ell_b} \eta_\phi \sum_{i=1}^{K+3}  \left[- {{e^{l}_{A}}} \left({m}_{\phi}^{2} - \left({{m_{H^{\pm}}}}\right)^{2} \right) \operatorname{C_{1}} {m}_{\ell_a}^{2} 
\right.\\ & \left. \qquad
+ {{e^{n}_{A}}} \left(\left({m}_{\phi}^{2}- \left({{m_{H^{\pm}}}}\right)^{2} \right)  \left( \operatorname{C_{0}} + \operatorname{C_{2}} \right)    \right) {{{m_n}}}_{i}^{2} \right] \Delta_{ii}^{ab},
\end{align}
\begin{align}\notag
A_L^{(16)}(n_i H^- H^+) & =  m_{\ell_a} \dfrac{16 \lambda_5 \rho_{\phi} m_{W}^{2} + g^{2} \left(\rho_{\phi} \mathcal{K}_{\phi} - \Delta_\phi \mathcal{Q}_\phi\right)}{2 g^{2} \sin{\left(2 \beta \right)}}\\ 
& \times \sum_{i=1}^{K+3} \left[ - \left({{e^{l}_{A}}}\right)^{2} \operatorname{C_{2}} {m}_{\ell_b}^{2} - {{e^{l}_{A}}} {{e^{n}_{A}}} \operatorname{C_{0}} {{{m_n}}}_{i}^{2} + \left({{e^{n}_{A}}}\right)^{2} \operatorname{C_{1}} {{{m_n}}}_{i}^{2}\right]  \Delta_{ii}^{ab},   \\ \notag
A_R^{(16)}(n_i H^- H^+) & =  {m}_{\ell_b} \dfrac{16 \lambda_5 \rho_{\phi} m_{W}^{2} + g^{2} \left(\rho_{\phi} \mathcal{K}_{\phi} - \Delta_\phi \mathcal{Q}_\phi \right)}{2 g^{2} \sin{\left(2 \beta \right)}}\\ 
& \times \sum_{i=1}^{K+3} \left[\left({{e^{l}_{A}}}\right)^{2} \operatorname{C_{1}} {m}_{\ell_a}^{2} - {{e^{l}_{A}}} {{e^{n}_{A}}} \operatorname{C_{0}} {{{m_n}}}_{i}^{2} - \left({{e^{n}_{A}}}\right)^{2} \operatorname{C_{2}} {{{m_n}}}_{i}^{2}\right]  \Delta_{ii}^{ab}.
\end{align}
\begin{align}
A_L^{(17)}(n_i H^{\pm}) & =   \dfrac{{m}_{\ell_a} {m}_{\ell_b}^2 }{{m}_{\ell_a}^{2} - {m}_{\ell_b}^{2}} c^l_\phi  \sum_{i=1}^{K+3}  \left(2 {{e^{l}_{A}}} {{e^{n}_{A}}} \operatorname{{{B^{(1)}_{0}}}} {{{m_n}}}_{i}^{2} - \left(\left({{e^{l}_{A}}}\right)^{2} {m}_{\ell_a}^{2} + \left({{e^{n}_{A}}}\right)^{2} {{{m_n}}}_{i}^{2}\right) \operatorname{{{B^{(1)}_{1}}}}\right) \Delta_{ii}^{ab},\\ \notag
A_R^{(17)}(n_i H^{\pm}) & =  \dfrac{ {m}_{\ell_b} }{{m}_{\ell_a}^{2} - {m}_{\ell_b}^{2}} c^l_\phi  \sum_{i=1}^{K+3} \left({{e^{l}_{A}}} {{e^{n}_{A}}} \left({m}_{\ell_a}^{2} + {m}_{\ell_b}^{2}\right) \operatorname{{{B^{(1)}_{0}}}} {{{m_n}}}_{i}^{2} 
\right.\\ & \left. \qquad \qquad
- \left(\left({{e^{l}_{A}}}\right)^{2} {m}_{\ell_b}^{2} + \left({{e^{n}_{A}}}\right)^{2} {{{m_n}}}_{i}^{2}\right) \operatorname{{{B^{(1)}_{1}}}} {m}_{\ell_a}^{2}\right)\Delta_{ii}^{ab},
\end{align}
\begin{align}\notag
A_L^{(18)}(H^{\pm} n_i) & = - \dfrac{ {m}_{\ell_a}}{{m}_{\ell_a}^{2} - {m}_{\ell_b}^{2}} c^l_\phi  \sum_{i=1}^{K+3} \left({{e^{l}_{A}}} {{e^{n}_{A}}} \left({m}_{\ell_a}^{2} + {m}_{\ell_b}^{2}\right) \operatorname{{{B^{(2)}_{0}}}} {{{m_n}}}_{i}^{2} 
\right.\\ & \left. \qquad \qquad
+ \left(\left({{e^{l}_{A}}}\right)^{2} {m}_{\ell_a}^{2} + \left({{e^{n}_{A}}}\right)^{2} {{{m_n}}}_{i}^{2}\right) \operatorname{{{B^{(2)}_{1}}}} {m}_{\ell_b}^{2}\right) \Delta_{ii}^{ab}, \\
A_R^{(18)}(H^{\pm} n_i) & = - \dfrac{ {m}_{\ell_a}^2 {m}_{\ell_b}}{{m}_{\ell_a}^{2} - {m}_{\ell_b}^{2}} c^l_\phi  \sum_{i=1}^{K+3}\left(2 {{e^{l}_{A}}} {{e^{n}_{A}}} \operatorname{{{B^{(2)}_{0}}}} {{{m_n}}}_{i}^{2} + \left(\left({{e^{l}_{A}}}\right)^{2} {m}_{\ell_b}^{2} + \left({{e^{n}_{A}}}\right)^{2} {{{m_n}}}_{i}^{2}\right) \operatorname{{{B^{(2)}_{1}}}}\right)\Delta_{ii}^{ab}.
\end{align}
%

\bibliography{biblioteca.bib}

\end{document}